\documentclass[preprint,11pt,review]{elsarticle}

\usepackage{lineno,hyperref}

\journal{Renewable and Sustainable Energy Reviews}

\usepackage{amsmath}
\usepackage{amsthm}
\usepackage[utf8]{inputenc}
\usepackage[english]{babel}

\newtheorem{proposition}{Proposition}

\usepackage[ruled,linesnumbered]{algorithm2e}
\usepackage{multirow}
\usepackage[bottom=1in,top=1in,left=1in,right=1in]{geometry}

\usepackage[table]{xcolor}
\usepackage{enumitem}

\usepackage{nomencl}
\usepackage{etoolbox}
\usepackage{bbding}
\usepackage{float}
\usepackage{bm}
\usepackage{amssymb}
\usepackage{indentfirst}  
\usepackage{graphicx}  
\usepackage{subfigure}
\usepackage{booktabs} 
\usepackage{threeparttable}   
\usepackage{algpseudocode}
\usepackage{arydshln}
\usepackage{color}
\usepackage{microtype}
\usepackage{cases}
\usepackage{enumerate}

\usepackage{makecell}
\usepackage{lscape}

\usepackage{amssymb}
\usepackage{nomencl}
\makenomenclature

\usepackage{etoolbox}
\renewcommand\nomgroup[1]{%
  \item[\bfseries
  \ifstrequal{#1}{B}{Sets}{%
  \ifstrequal{#1}{F}{First-stage decision variables}{%
  \ifstrequal{#1}{G}{Second-stage decision variables}{%
  \ifstrequal{#1}{C}{Constant parameters}{%
  \ifstrequal{#1}{A}{Indices}{%
  \ifstrequal{#1}{R}{Random variables}{}}}}}}%
]}

\usepackage{tcolorbox}

\begin{document}
\nomenclature[A]{$t$}{Index of time period}
\nomenclature[A]{$i$}{Index of distribution node}
\nomenclature[A]{$(i,j)$}{Index of power line from node $i$ to node $j$ (directed)}%
\nomenclature[A]{$h$}{Index of line hardening measure}
\nomenclature[A]{$zn$}{Index of geographical zone}
\nomenclature[A]{$tr$}{Index of hurricane propagation track scenario}

\nomenclature[B]{$\mathcal{T}$}{Set of time periods}
\nomenclature[B]{$\Omega^{\rm L},\Omega^{\rm L}_{zn}$}{Set of distribution lines, and distribution lines in zone $zn$}
\nomenclature[B]{$\mathcal{H}$}{Set of line hardening measures}
\nomenclature[B]{$\Omega^{\rm N},\Omega^{\rm S},\Omega^{\rm Z}$}{Set of nodes, substations and geographical zones}
\nomenclature[B]{$\mathcal{N}_{sc}$}{Set of hurricane track scenarios}
\nomenclature[B]{$\Omega^{\rm SW}$}{Set of lines with installed switches}

\nomenclature[F]{$y^h_{ij}$}{Binary variable indicating the hardening decision; equals 1 if line $(i,j)$ is hardened by measure $h$, 0 otherwise}
\nomenclature[F]{$\boldsymbol{y}$}{Vector including all hardening decision variables $y^h_{ij}$}
\nomenclature[F]{$s_{ij}$}{Binary variable indicating the switching state; equals 1 if line $(i,j)$ is switched, 0 otherwise}
\nomenclature[F]{$x^G_i$}{Binary variable indicating the DER allocation decision; equals 1 if a DG is allocated at node $i$, 0 otherwise}
\nomenclature[F]{$x^S_{ij}$}{Binary variable indicating the switching state of line $(i,j)$; equals 1 if the switch is on, 0 otherwise}
\nomenclature[F]{$f_{ij}$}{Fictitious flow of line $(i,j)$}
\nomenclature[F]{$\kappa_{i}$}{Binary variable indicating the root node; equal 1 if a bus $i$ is chosen as a root node}

\nomenclature[C]{$N^{\rm L}_{zn,t}$}{Maximum number of affected power lines during the contingency in zone $zn$}
\nomenclature[C]{$N^{\rm C}$}{Number of conductor wires between two adjacent poles}
\nomenclature[C]{$N^{\rm P}$}{Number of distribution poles supporting each lines}
\nomenclature[C]{$\gamma_i$}{Priority weight of demand at node $i$}
\nomenclature[C]{$P^{\rm D}_{i,t},Q^{\rm D}_{i,t}$}{Active and reactive power load at node $i$ in time $t$}
\nomenclature[C]{$\hat{P}^{\rm L}_{ij},\hat{Q}^{\rm L}_{ij}$}{Upper/lower bound of active power capacity through line $(i,j)$}
\nomenclature[C]{$R_{ij},X_{ij}$}{Resistance and reactance of distribution line $(i,j)$}
\nomenclature[C]{$RU_{i},RD_{i}$}{Ramping-up/ramping-down rate of DER at node $i$}
\nomenclature[C]{$x^{\rm sub}_i$}{Binary parameter indicating the substation location; equals 1 if the substation is at node $i$}
\nomenclature[C]{$\theta_{tr}$}{Probability of hurricane track scenario $tr$}
\nomenclature[C]{$C^h_{ij}$}{Hardening cost of line $(i,j)$ under $h$ hardening measure}
\nomenclature[C]{$C^G,C^S$}{Cost of DER allocation and switch placement}
\nomenclature[C]{$B^G,B^L$}{Available budget for DG allocation and line hardening}
\nomenclature[C]{$\hat{U}^{sqr}_{i},\check{U}^{sqr}_{i}$}{Maximum/minimum squared voltage of node $i$}
\nomenclature[C]{$\hat{P}_{i}^G,\hat{Q}_{i}^G$}{Maximum active/reactive DER output at node $i$}
\nomenclature[C]{$x^{S}_{ij}$}{Binary parameter indicating the switch location; equals 1 if the line $(i,j)$ is installed with switch}

\nomenclature[R]{$u_{ij,tr,t}$}{Bernoulli random state variable; equals 0 if line $(i,j)$ is affected in period $t$ in scenario $tr$, 1 otherwise.}
\nomenclature[R]{${v}_{tr,zn,t}$}{Wind speed of the hurricane at zone $zn$ in period $t$ under hurricane track scenario $tr$}

\nomenclature[G]{$\sigma_{tr,i,t}$}{Load shedding coefficient of node $i$ in time $t$ under hurricane track $tr$}
\nomenclature[G]{$U^{\rm sqr}_{tr,i,t}$}{Squared voltage magnitude of  node $i$ in time $t$ under hurricane track $tr$}
\nomenclature[G]{$p^{\rm G}_{tr,i,t},q^{\rm G}_{tr,i,t}$}{Active/reactive power output of DER at node $i$ in time $t$ under hurricane track $tr$}
\nomenclature[G]{$p^{\rm sub}_{tr,i,t},q^{\rm sub}_{tr,i,t}$}{Active/reactive power output of substation at node $i$ in time $t$ under hurricane track $tr$}
\nomenclature[G]{$p^{\rm L}_{tr,ij,t},q^{\rm L}_{tr,ij,t}$}{Active/reactive power on distribution line $(i,j)$ in time $t$ under hurricane track $tr$}
\setlength{\nomitemsep}{-\parsep}

\begin{frontmatter}

{\title{\LARGE A Distributionally Robust Resilience Enhancement Strategy for Distribution Networks Considering Decision-Dependent Contingencies}}




\author[mymainaddress]{Yujia Li}
\ead{yjli@eee.hku.hk}
\author[mysecondaryaddress]{Shunbo Lei}
\ead{leishunbo@cuhk.edu.cn}
\author[mythirdaddress]{Wei Sun}
\ead{sun@ucf.edu}
\author[mymainaddress]{Chenxi Hu}
\ead{cxhu@eee.hku.hk}
\author[mymainaddress]{Yunhe Hou}
\ead{yhhou@eee.hku.hk}
\address[mymainaddress]{Department of Electrical and Electronic Engineering, The University of Hong Kong, Hong Kong 999077, China}
\address[mysecondaryaddress]{School of Science and Engineering, The Chinese University of Hong Kong, Shenzhen, 
Guangdong 518172, China}
\address[mythirdaddress]{Department of Electrical and Computer Engineering, University of Central Florida, Orlando, FL 32816, USA}

\vspace{-3ex}
\begin{abstract}
When performing the resilience enhancement for distribution networks, there are two obstacles to reliably model the uncertain contingencies: 1) decision-dependent uncertainty (DDU) due to various line hardening decisions, and 2) distributional ambiguity due to limited outage information during extreme weather events (EWEs). To address these two challenges, this paper develops scenario-wise decision-dependent ambiguity sets (SWDD-ASs), where the DDU and distributional ambiguity inherent in EWE-induced contingencies are simultaneously captured for each possible EWE scenario. Then, a two-stage trilevel decision-dependent distributionally robust resilient enhancement (DD-DRRE) model is formulated, whose outputs include the optimal line hardening, distributed generation (DG) allocation, and proactive network reconfiguration strategy under the worst-case distributions in SWDD-ASs. Subsequently, the DD-DRRE model is equivalently recast to a mixed-integer linear programming (MILP)-based master problem and multiple scenario-wise subproblems, facilitating the adoption of a customized column-and-constraint generation (C\&CG) algorithm. Finally, case studies demonstrate a remarkable improvement in the out-of-sample performance of our model, compared to its prevailing stochastic and robust counterparts. Moreover, the potential values of incorporating the ambiguity and distributional information are quantitatively estimated, providing a useful reference for planners with different budgets and risk-aversion levels.
\\
\textbf{Keywords}: 
Decision-dependent uncertainty,
distribution network,
distributionally robust optimization,
line hardening,
network reconfiguration,
resilience enhancement.
\end{abstract}
\end{frontmatter} 


{\setlength{\parsep}{-1pt}
\setlength{\parskip}{-1pt}
\small
\printnomenclature[0.9in]
}

\section{Introduction}

Nowadays, ongoing climate change is placing an excessive strain on the environment, giving rise to the ever-increasing frequency and intensity of extreme weather events (EWEs). Power systems, as one of the most widespread and critical infrastructures, have been seriously affected by those EWEs over the past decades. According to 
\citet{DOEE}, 58\% of U.S. power outages arise from EWEs such as hurricanes, winter storms and floods, leading to an annual economic loss of \$18-33 billion. In this context, power system resilience, defined as the capability of a system to anticipate, absorb and recover from severe events in a timely and successful manner \citep{field2012managing}, becomes one of the most crucial aspects in the design of electrical grids. In addition, given that nearly 90\% of storm-related outages are reported to happen within the radial distribution grids \citep{DOEE}, the significance of resilience investments on the distribution side should be further underlined. 

Conducting proactive physical hardening has been a vital part of the long-term design of resilient distribution systems. 
To defend against the possible EWEs in the upcoming years, several measures are commonly employed, including reinforcing overhead structures \citep{9209142}, undergrounding cables \citep{trakas2021strengthening}, and managing nearby vegetation \citep{1046899}. Moreover, recent advancements in active distribution networks (ADNs) allow active units, e.g., distributed energy resources (DERs), remotely controlled switches (RCSs) and responsive loads, to actively participate in the resilient operation \cite{taheri2020distribution}, \cite{8080265}. 
Given the limited budget in the real-world planning, it is critical to identify the most vulnerable components and best locations for active resources.

Accurately modeling the uncertain contingencies under EWEs is the prerequisite for avoiding suboptimal enhancement strategies. 
While the physical hardening problems of ADNs have been extensively investigated (see, e.g., \cite{9209142},  \cite{trakas2021strengthening},  \cite{1046899}, \cite{taheri2020distribution}, \cite{8080265}), two fundamental issues with contingency quantification remain unresolved: 1) probabilistic contingency models are always subject to misspecification, due to the scarcity of historical data on EWEs as well as the insufficient knowledge of interactions between weather and infrastructure \citep{hughes2021damage}; 2) the \emph{decision-dependent uncertainty (DDU)} inherent in contingencies has not been adequately characterized in the existing literature.

\begin{figure}
  \centering
  \includegraphics[width=13.5cm,height=5.2cm]{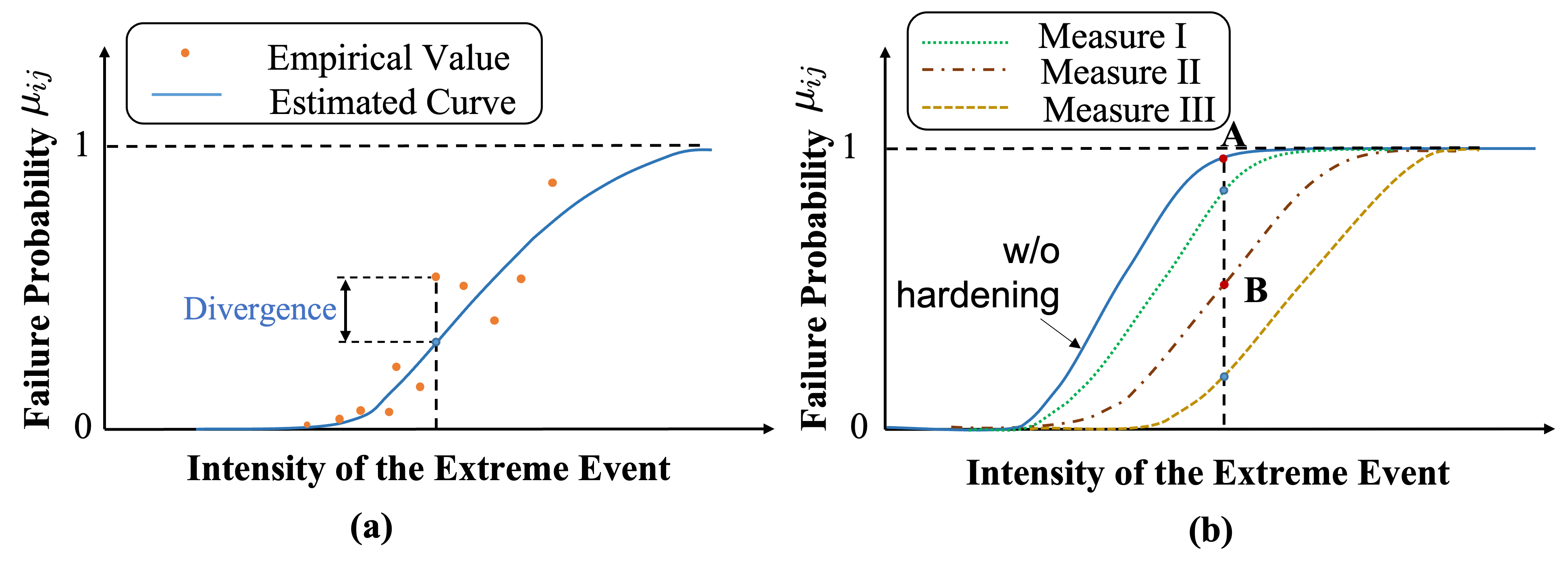}\\
  \caption{(a) Empirical failure probabilities and estimated fragility curve of line $(i,j)$; (b) Decision-dependent fragility functions of line $(i,j)$ without hardening and under different hardening measures I, II and III}
  \label{motivation}
\end{figure}

Concerning the first issue, while multiple studies have applied data-driven approaches \citep{wanik2018case}, physical models \citep{yuan2018resilience} and their hybridization \citep{hughes2021damage} to forecast contingencies, 
many evidences exhibit their limited accuracy. 
Figure \ref{motivation}(a) illustrates the potential divergences between the empirical failure probabilities and the estimated fragility curve of a distribution line \citep{dunn2018fragility}, 
which mainly arise from the unavoidable model bias and inherent variance of the component's fragility. 
Nonetheless, both of the most prevalent frameworks dealing with contingency uncertainty, namely the stochastic programming (SP) and robust optimization (RO), are incapable of integrating this estimation ambiguity. 
Specifically, the SP-based models optimistically adopt the estimated fragility curves by entirely trusting its accuracy (see, e.g., \cite{8329529}, \cite{9497740}), while RO-based models pessimistically utilized the N-k uncertainty sets by giving up the accessible distribution information  (see, e.g., \cite{9521770}, \cite{7381672}). 
Unfortunately, these two assumptions might result in suboptimal solutions as a result of their excessive optimism and pessimism. 

Advancements in distributionally robust optimization (DRO) offer an alternative for solving problems with ambiguous probabilistic models. It has already been applied in several resilience-related problems, including microgrid formation \citep{9477426}, network configuration \citep{8998220}, and resilient operation \citep{9316222}. DRO-based models generate decisions based on worst probability distribution in an ambiguity set, 
yielding more robust and realistic results (compared to SP and RO, respectively). 
However, former studies only involve static ambiguity sets for \emph{decision-independent uncertainties (DIU)}, leaving DDU-related contingencies unexplored.

Concerning the second issue, extra efforts are required to tractably capture the interdependence between random contingencies and hardening decisions. In addition to exogenous factors such as the intensity of the EWEs and the aging states of the physical infrastructure, realization of contingencies can be endogenously controlled by purposeful hardening, so the associated uncertainty should be classified as DDU.
The four empirical failure probability functions under different hardening decisions depicted in Figure \ref{motivation} (b) illustrate this point \citep{yuan2018resilience}. 

Currently, there are only several related studies establishing explicit formulations for this DDU. 
In \citet{8329529}, the DDU of contingencies are incorporated in an SP-based framework to construct a hardening strategy. A trilevel RO-based hardening model is proposed in \citet{7514755}, where coupling between first-level hardening decision and uncertain contingencies is formulated in a uncertain set. However, the additional computational complexity imposed by DDU always necessitate the scenario reduction techniques \citep{8329529} or heuristic algorithm \citep{7514755}. 
Furthermore, they both overlook the potential ambiguity in contingency estimation. On the other hand, in the research community of operations research, DRO problems with DDU have recently received increased attention. For example, both moment and distance-based DRO problems with DDU have been studied (see, e.g., \cite{noyan2018distributionally},  \cite{luo2020distributionally}), but finding effective algorithms for solving their non-convex reformulations remains a pending issue. 
To solve the DRO-based nurse staffing problem with DDU, \citeauthor{ryu2019nurse} developed a mixed-integer linear programming (MILP) reformulation and employed a separation algorithm.
In \cite{basciftci2021distributionally}, a DRO-based facility-location problem, in which moments of demand are dependent on location decisions, is reformulated into an exact MILP. Inspired by these pioneering studies, this paper tries to fill the research gap by making the following contributions:

\begin{itemize}
    \item[1)] We propose novel scenario-wise decision-dependent ambiguity sets (SWDD-ASs) to simultaneously incorporate the ambiguity and DDU of EWE-induced contingencies. 
    Moreover, two DIUs, i.e., the dynamic track and intensity of EWEs are mapped to the SWDD-ASs by means of scenarios and confidence intervals, respectively. 
    
    \item[2)] Based on the worst-case distributions identified in SWDD-ASs, a two-stage trilevel decision-dependent distributionally robust resilience enhancement (DD-DRRE) model is constructed to minimize the expected weighted load shedding (EWLS). 
    The final outputs are the optimal measures to harden overhead structures and allocate DERs, as well as to proactively reconfigure the network. 
    
    \item[3)] Through exploring its strong duality properties and equivalent linearization techniques, the proposed two-stage trilevel DD-DRRE model is recast to an MILP-based master problem and multiple MILP-based scenario-wise subproblems. A customized column-and-constraint generation (C\&CG) algorithm is applied to effectively solve it.  
    
    \item[4)] 
    Extensive in-sample and out-of-sample evaluations are conducted to compare our method to its SP- and RO-based counterparts. 
    Additionally, the explicit computation of the worst-case distributions and two novel metrics, namely the \emph{value of distributional ambiguity (VoDA)} and the \emph{value of moment information (VoMI)}, 
    can serve as effective evaluation tools for planners to select a preferred strategy under different budgets and risk-aversion levels.  
\end{itemize}

The remainder of this paper is organized as follows. We first introduce SWDD-ASs in Section \ref{Section 2}. Then, we provide a detailed formulation of the two-stage trilevel DD-DRRE model in Section \ref{Section 3}. In section \ref{Section 4}, the tractable reformulation is derived for the DD-DRRE model. Evaluation methods and two metrics are introduced in Section \ref{Section 5}. In section \ref{Section 6}, numerical tests are conducted to demonstrate the efficiency of the proposed approach. Finally, Section \ref{Section 7} concludes our study.

\section{Scenario-wise Decision-dependent Ambiguity Sets for Contingencies}\label{Section 2}

\subsection{Assumptions}
For narrowing down the focus without loss of generality, we make the following assumptions throughout the paper:
\begin{itemize}
    \item[1)] We choose hurricanes as the representative EWE for their prevalence and serious consequences \citep{DOEE}. Therefore, hardening measures are focused on overhead structures. 
    \item[2)] The entire network can be divided into several zones.
    Due to the small expanse of each zone, the meteorological characteristics within each zone are assumed identical \citep{9209142}.
    \item[3)] Following \citet{8329529} and \citet{9316222}, the failure probabilities of different components are assumed to be independent.
    \item[4)] We assume that system planners have access to the prediction on hurricane pro\-pa\-ga\-tion tr\-acks and in\-ten\-si\-ties for the studied time horizon, either estimated from historical data or fo\-re\-cast by the hurricane center \cite{trakas2021strengthening}, \cite{javanbakht2014risk}.
\end{itemize}
\subsection{Scenario-wise Decision-dependent Ambiguity Sets of Contingency Distributions}
Considering the distributional ambiguity of contingencies, we focus on the worst-case distributions within SWDD-ASs and their corresponding EWLS. 
For illustrating the fundamental rationale, we first present a monolithic formulation for SWDD-ASs and leaving some details in the next subsection.

In the context of long-term resilience enhancement, the uncertainty associated with the outage state during a EWE are primarily influenced by three factors: 1) the hardening decision, which is an endogenous factor that planners can manipulate; 2) spatiotemporal track of hurricane and 3) hurricane intensity, which are exogenous factors that are completely dependent on nature.
For the second factor, we suppose that $|\mathcal{N}_{sc}|$ discrete scenarios are estimated to represent the possible tracks, while the uncertain hurricane intensity is captured by a confidence interval for wind speed \citep{javanbakht2014risk}, i.e., $[\check{v}_{tr,zn,t},\hat{v}_{tr,zn,t}]$. Then, for each track scenario $tr\in\mathcal{N}_{sc}$, the true contingency distribution 
is assumed to be within the moment-based SWDD-AS as follows: 
\begin{small}
\begin{subequations}
\begin{align}
&\mathcal{A}_{tr}(\boldsymbol{y})=\
\big\{\mathbb{P}_{tr}\in \mathcal{P}\big(\mathcal{U}_{tr}\big)\big|\forall (i,j)\in\Omega^{\rm L}_{zn},t\in\mathcal{T},zn\in\Omega^{\rm Z}:\notag
\\
&\tilde{f}_{ij}(\check{v}_{tr,zn,t},\boldsymbol{y}_{ij})\le \mathbb{E}_{\mathbb{P}_{tr}}[1\!-\!{u}_{ij,tr,t}]
\le\tilde{f}_{ij}(\hat{v}_{tr,zn,t},\boldsymbol{y}_{ij})\big\}\tag{1}\label{ambiguity set-wind speed}
\end{align}\label{AS}
\end{subequations}
\end{small}
where the supporting set $\mathcal{U}_{tr}$ is defined as:
\begin{small}
\begin{subequations}
\begin{align}
    &\mathcal{U}_{tr}=
    \big\{{u}_{ij,tr,t}\big|\!\!\!\!
    \sum_{(i,j)\in\Omega^{\rm L}_{zn}}\!\!\!(1\!-\!u_{ij,tr,t})\le N^{\rm L}_{zn,t},\forall t\!\in\!\mathcal{T},zn\!\in\!\Omega^{\rm Z}\label{max-line-outage}
    \\
    &\sum_{t=1}^{\lambda_{ij}^{\rm M}}\!(1\!\!-\!\!u_{ij,tr,t+\lambda})\!\ge\! \lambda_{ij}^{\rm M} (u_{ij,tr,t}\!\!-\!u_{ij,tr,t+1}),
     \forall t\!\in\!\mathcal{T}\!,\! (i,j)\!\in\!\Omega^{\rm L}\!\big\}
     \label{sequential-failure}
\end{align}
\label{supporting set}
\end{subequations}
\end{small}
The $\mathcal{P}(\mathcal{U}_{tr})$ in \eqref{AS} contains all distributions in the sigma-field of  $\mathcal{U}_{tr}$, and $\mathbb{E}_{\mathbb{P}_{tr}}[\cdot]$ is the expectation operator under distribution $\mathbb{P}_{tr}$. The constraints in set \eqref{ambiguity set-wind speed} imply that the marginal failure probability of each distribution line $(i,j)$ 
is within the interval $[\tilde{f}_{ij}(\check{v}_{tr,zn,t},\boldsymbol{y}_{ij}),\tilde{f}_{ij}(\hat{v}_{tr,zn,t},\boldsymbol{y}_{ij})]$. 
$\tilde{f}_{ij}(\cdot)$ is the decision-dependent 
failure probability function of line $(i,j)$, whose value is parametrized by the exogenous factor, i.e., wind speed $v_{tr,zn,t}$, and the endogenous factors, i.e., hardening decisions $\boldsymbol{y_{ij}}=[y^h_{ij},h\in\mathcal{H}]^{\rm T}$.
The overhead tilde in $\tilde{f}_{ij}(\cdot)$ indicates its ambiguity due to misspecified estimation. 

Moreover, $\mathcal{U}_{tr}$ defined in \eqref{supporting set} specifies the set of all plausible contingency scenarios. 
Constraints \eqref{max-line-outage} consider the number of concurrently outaged lines at zone $zn$, which is bounded by $N^L_{zn}$ and can be calibrated based on reliability analyses \cite{8998220}.
Constraints \eqref{sequential-failure} model the minimum restoration time $\lambda_{ij}^{\rm M}$. 

\subsection{Decision-dependent Ambiguous Failure Probability Functions}
To derive a closed-form formulation for 
$\tilde{f}_{ij}(v_{tr,zn,t},\boldsymbol{y}_{ij})$, fragility analyses toward overhead structures are necessary. This subsection 
addresses the challenges in the fragility estimation from three aspects: 1) decision-dependence, 2) stochasticity of hurricane intensity, and 3) estimation errors. 

\subsubsection{Decision-dependent Empirical Fragility Curves}
\begin{table}[]\footnotesize
\centering
\caption{Line Hardening Measures}
\label{Measures}
\vspace{1ex}
\begin{tabular}{ccccc}
\hline\hline
$h$ & \textbf{Measure}                   & \begin{tabular}[c]{@{}c@{}}\textbf{Decision}\\ \textbf{Variable}\end{tabular} & \begin{tabular}[c]{@{}c@{}}\textbf{Failure}\\ \textbf{Probability}\end{tabular} & \begin{tabular}[c]{@{}c@{}}\textbf{Estimation} \\ \textbf{Error}\end{tabular} \\ \hline
I   & Vegetation Management          &   $y^{\rm I}_{ij}$                &      $\mu_{ij}^{\rm{I}}(v)$   &
$\varepsilon_{ij}^{\rm{I}}(v)$\\
II  & Pole Replacement     &   $y^{\rm II}_{ij} $               &      $\mu_{ij}^{\rm{II}}(v)$   &
$\varepsilon_{ij}^{\rm{II}}(v)$\\
III & Combination of I and II &   $y^{\rm III}_{ij}$                &      $\mu_{ij}^{\rm{III}}(v)$   &
$\varepsilon_{ij}^{\rm{III}}(v)$\\ \hline\hline
\end{tabular}
\vspace{-2ex}
\end{table}
Fragility curves, which link the failure probability of network components with the intensity of natural hazards, are widely used in failure evaluation of structural systems. 
During a hurricane, the wind can destroy the overhead structure in either direct or indirect ways, i.e., directly blowing down the poles/wires, or blowing down the nearby trees to indirectly destroy wires. In this regard, three targeted hardening measures are herein considered as listed in Table \ref{Measures} \citep{8329529}. 
Therefore, empirical fragility curves can be derived for poles and conductor wires with regard to wind speed $v$ \citep{ouyang2014multi}:
\begin{align}
    &\mu^{\rm P}_{ij,k}(v,\boldsymbol{y}_{ij})=\phi\big( \frac{{\rm ln}({v}/{m_{ij,k}(\boldsymbol{y}_{ij})})}{\sigma_{ij,k}(\boldsymbol{y}_{ij})}\big)\qquad\qquad \forall k\in\Omega_{ij}^P \label{fragility-pole}
    \\
   & \mu^{\rm C}_{ij,l}(v,\boldsymbol{y}_{ij})\!=\!{\rm max} \big(\mu^{\rm C,\!W}_{ij,l}\!(v),\chi_{ij,l}\!(v)\mu^{\rm C,\!VE}_{ij,l}\!(v,\boldsymbol{y}_{ij})\big),
    \forall l\!\in\!\Omega_{ij}^{C}\label{veg}
\end{align}
\indent \indent For poles, the lognormal cumulative distribution function $\phi(\cdot)$ are used to present their empirical fragility, as \eqref{fragility-pole} shows \cite{salman2015evaluating}.
$m_{ij,k}(\boldsymbol{y}_{ij})$ and $\sigma_{ij,k}(\boldsymbol{y}_{ij})$ are parameters whose values can be altered by replacing poles. 
On the other hand, equations \eqref{veg} define the fragility curves of conductor wires, where
$\mu^{\rm C,W}_{ij,l}(v)$ and $\chi_{ij,l}(v)$ are the direct wind-induced and the fallen tree-induced failure probability of conductor $l$ \cite{ouyang2014multi}.
$\mu^{\rm C,VE}_{ij,l}(v)$ is the falling probability of nearby vegetation, which can be reshaped to be 0 by managing the nearby vegetation.

\begin{figure}
  \centering
  \includegraphics[width=7cm,height=5.5cm]{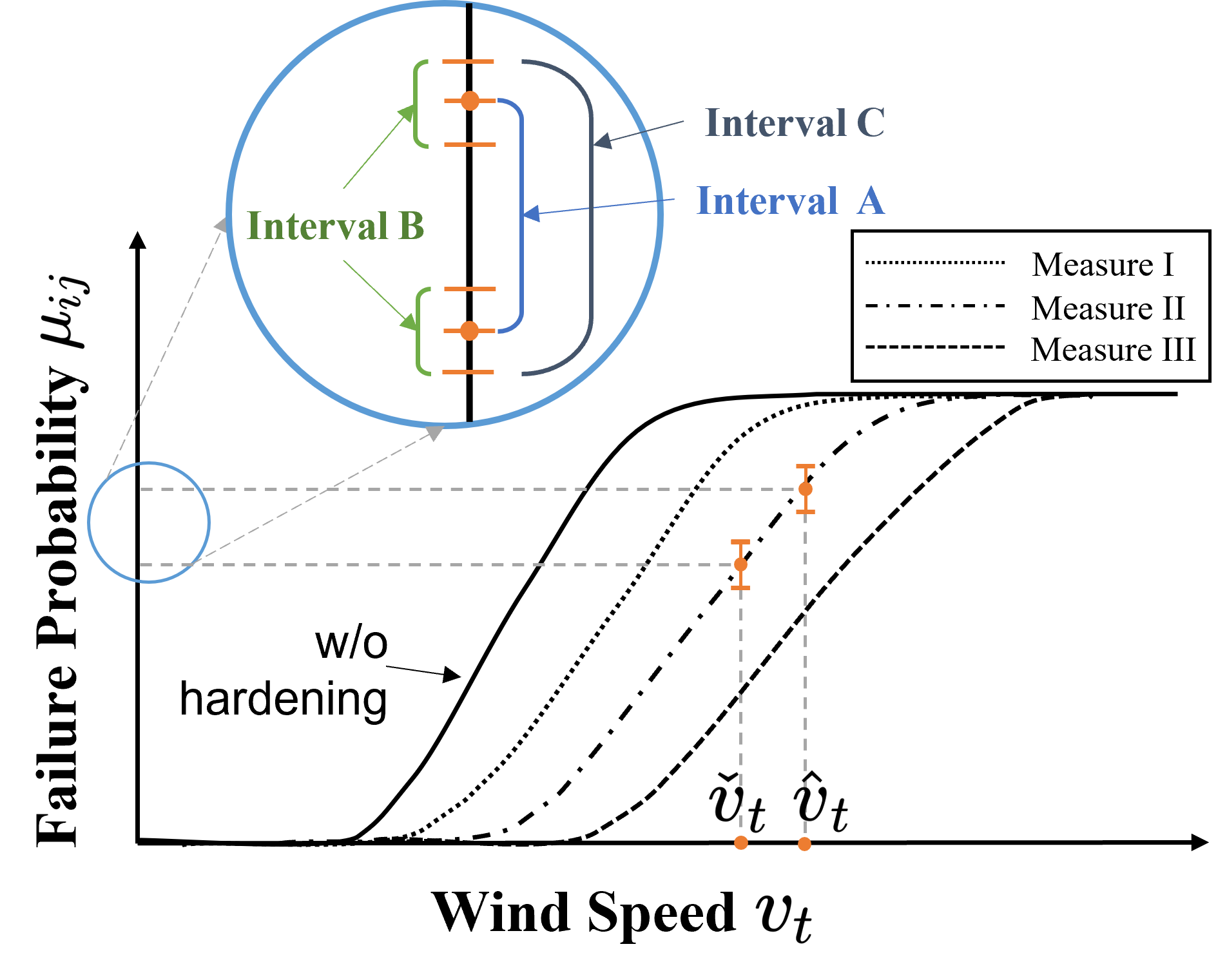}\\
  \caption{Decision-dependent failure probability functions of line $(i,j)$ under different line hardening measures 
  }
  \label{4Curves}
\end{figure}

Then, as the breakdown of a single pole or conductor results in the disconnection of the entire line, the empirical fragility curves for the distribution line $(i,j)$ is derived as below:

\vspace{-2ex}
\begin{small}
\begin{equation}
\begin{aligned}
    \mu_{ij}(v,\boldsymbol{y}_{ij})=
    1\!-\!\prod_{k=1}^{N_{ij}^{\rm P}}\big(1\!-\!\mu^{\rm P}_{ij,k}(v,\boldsymbol{y}_{ij})\big)\prod_{l=1}^{N_{ij}^{\rm C}}\big(1\!-\!\mu^{\rm C}_{ij,l}(v,\boldsymbol{y}_{ij})\big)
    \end{aligned}
    \label{outage probability}
\end{equation}
\end{small}
\!\!To intuitively show the impacts of $\boldsymbol{y}_{ij}$ and $v$, four independent empirical curves computed through \eqref{fragility-pole}-\eqref{outage probability} with different $\boldsymbol{y}_{ij}$ are depicted in Fig. \ref{4Curves}. 
The solid curve corresponds to the case where no hardening measure is conducted on an aging line $(i,j)$ (i.e., $\boldsymbol{y}_{ij}=\boldsymbol{0}$), denoted as $\mu_{ij}^{0}(v)$. The rest of curves are denoted as $\mu_{ij}^{h}(v)$. By utilizing $\mu^0_{ij}(v)$ and $\mu^h_{ij}(v)$, the empirical fragility curves \eqref{outage probability} can be further rewritten as the following piecewise linear functions in binary variables $\boldsymbol{y}_{ij}$: 
\begin{subequations}
\begin{align}
    {\mu}_{ij}(v,\boldsymbol{y}_{ij})\! &=\! {\mu}^0_{ij}(v)\!+\!\sum_{h\in\mathcal{H}}\Delta{\mu}^h_{ij,t}(v)y^h_{ij}\label{piecewise linear}
    \\
    \Delta{\mu}^h_{ij,t}(v)\!&=\!\mu^h_{ij}({v})\!-\!{\mu}^0_{ij}(v),\qquad
    \forall{(i,j)\in\Omega^L,t\in\mathcal{T}}
    \label{Delta}
\end{align}\label{empirical-fragility}
\end{subequations}
\!\!\!\! where $\Delta{\mu}^h_{ij}(v_{zn,t})$ in \eqref{Delta} is the reduction in empirical failure probability after implementing the hardening measure $h$ on line $(i,j)$, which is a negative value.

\subsubsection{Decision-dependent Intensity-related Interval} 
While the forecasting toward hurricane track and intensity improves greatly in previous years, evidences still show comparability large yearly mean absolute errors due to the complexity of air-sea interactions and thermodynamic process \citep{shimada2018further}, especially the long-term forecasting of hurricane intensity \citep{trabing2020understanding}. Therefore, this DIU should be incorporated to maintain the robustness of the SWDD-ASs. Since distinct wind speeds will lead to different line failure probabilities, it is critical to derive the speed range at each time period $t$ under each possible track $tr$. 
According to the dynamic gradient wind field of a hurricane, $v_{tr,zn,t}$ at zone $zn$ on track $tr$ can be formulated as a function of the distance $d_{zn,t}$ to the hurricane eye \citep{javanbakht2014risk}. The detailed formulation is as below:
\begin{equation}
\begin{aligned}
v_{tr,zn,t} 
=&
\begin{cases}K_{tr}^{\rm C}V_{tr,t}^{\rm M}\big(1-{\rm exp}\big[\frac{1}{R_{tr,t}^{\rm M}}{\rm ln}(\frac{K_{tr}^{\rm C}}{K_{tr}^{\rm C}-1})d_{zn,t}], 0\le d_{zn,t}\le\! R^{\rm M}_{tr,t}\ &
  \\
  V_{tr,t}^{\rm M}{\rm exp}\big[\!-\!\big(\frac{{\rm ln}K^{\rm B}_{tr}}{R_{tr,t}^{\rm B}-R_{tr,t}^{\rm M}}\big)(d_{zn,t}\!-\!R_{tr,t}^{\rm M})\big], R^{\rm M}_{tr,t}\!\!\le\!\! d_{zn,t}\!\!\le\! \!R^{\rm B}_{tr,t}\ &
  \\
  0, \qquad \qquad\qquad\qquad \qquad\qquad\qquad \qquad d_{zn,t} \!>\! R^{\rm B}_{tr,t}
\end{cases}\\
& \qquad \qquad \qquad \forall tr\in\mathcal{N}_{sc}, \forall zn\in\Omega^Z,t\in\mathcal{T} \label{hurricane model}
\end{aligned}
\end{equation}

\noindent \!\! where parameters $K^{\rm C}_{tr}$ and $K_{tr}^{\rm B}$ represent the hurricane translation speed and boundary, which are assumed constant under each track scenario $tr$. 
${V}^{\rm M}_{tr,t}$ represents the maximum sustained wind speed of a hurricane, and ${R}^{\rm M}_{tr,t}$ and ${R}^{\rm B}_{tr,t}$ are the radius of the maximum wind speed and the radius of the affected area. 
To obtain a reliable interval of wind speed, we follow the method in \cite{javanbakht2014risk} to fit the probability distribution functions (PDFs) of parameters $\{\tilde{V}^M_{tr,t},\tilde{R}^B_{tr,t},\tilde{R}^M_{tr,t}\}$ using historical hurricane data, based on which PDFs of wind speed is computed through \eqref{hurricane model}. Then, by assigning a confidence level (e.g., 3$\sigma$ criterion if Gaussian distribution is adopted), the confidence interval for wind speed, i.e., 
$[\check{v}_{tr,zn,t}(\boldsymbol{y}_{ij}),\hat{v}_{tr,zn,t}(\boldsymbol{y}_{ij})]$ is derived. 
Thus, an interval $[\check{\mu}_{ij,tr,t},\hat{\mu}_{ij,tr,t}]$ for empirical failure probability induced by the randomness of hurricane intensity is calculated: 
\begin{subequations}
\begin{align}
    \check{\mu}_{ij,tr,t}(\boldsymbol{y}_{ij})\! &=\! {\mu}^0_{ij,t}(\check v_{tr,zn,t})\!+\!\sum_{h\in\mathcal{H}}\Delta{\mu}^h_{ij,t}(\check v_{tr,zn,t})y^h_{ij},\label{lower-mu}
    \\
    \hat{\mu}_{ij,tr,t}(\boldsymbol{y}_{ij})\! &=\! {\mu}^0_{ij,t}(\hat v_{tr,zn,t})\!+\!\sum_{h\in\mathcal{H}}\Delta\hat{\mu}^h_{ij,t}(\hat v_{tr,zn,t})y^h_{ij}\label{upper-mu}
    \\
    \forall (i,j)\in&\ \Omega_{zn}^L, tr\in\mathcal{N}_{sc}, zn\in\Omega^Z,t\in\mathcal{T},h\in\mathcal{H}\notag
\end{align}\label{upper/lower probability}
\end{subequations}
\noindent The lower and upper limits \eqref{lower-mu} and \eqref{upper-mu} for the interval are both piecewise linear functions in $\boldsymbol{y}_{ij}$ with $h+1$ independent coefficients. We name this interval as \emph{intensity-related interval} and 
\emph{Interval A} in Fig. \ref{4Curves} illustrates it under $y_{ij}^{\rm II}=1$.

\subsubsection{Decision-dependent Estimation Error-related Interval}
Besides the uncertain intensity, the ambiguity of empirical fragility curves \eqref{fragility-pole}-\eqref{veg} should be accounted for yielding more realistic results. The ambiguity of empirical fragility curves mainly result from two sources: 1) misspecified estimation model, and 2) inherent uncertainty of the failure probability itself. 
According to \cite{dunn2018fragility}, both the 
degrees of the two uncertainty sources increase when wind speed grows. 
Due to the main focus of this section is to construct SWDD-ASs, the pertinent details on fragility curve fitting are not included. 
Therefore, based on statistical results under different hardening decisions $\boldsymbol{y}_{ij}$, a disturbance term $\varepsilon_{ij} ({v}_{tr,zn,t},\boldsymbol{y}_{ij})$
is added to empirical failure probability $\mu_{ij,t}(v_{tr,zn,t},\boldsymbol{y}_{ij})$ to reflect the underlying ambiguity. Considering the binary of $\boldsymbol{y}_{ij}$, 
the holistic upper and lower bounds for error terms can be written as the following piecewise formulation:
\vspace{-4ex}
\begin{subequations}
\begin{align}
    &\check{\varepsilon}_{ij,t}(\boldsymbol{y}_{ij})\!=\!{\varepsilon}^0_{ij,t}(\check{v}_{tr,zn,t}) + \sum_{h\in\mathcal{H}}\Delta{\varepsilon}_{ij,t}^h(\check{v}_{tr,zn,t}) y^h_{ij}\label{8a}
    \\
    &\hat{\varepsilon}_{ij,t}(\boldsymbol{y}_{ij})\!=\!{\varepsilon}^0_{ij,t}(\hat{v}_{tr,zn,t}) + \sum_{h\in\mathcal{H}}\Delta{\varepsilon}_{ij,t}^h(\hat{v}_{tr,zn,t}) y^h_{ij}\label{8b}
    \\
    &\Delta{\varepsilon}^h_{ij}(v_{tr,zn,t})\!=\!\varepsilon^h_{ij}({v}_{tr,zn,t})-{\varepsilon}^0_{ij}({v}_{tr,zn,t})
    \label{Delta-error}
    \\
    &\forall (i,j)\in\Omega_{zn}^L,tr\in\mathcal{N}_{sc}, zn\in\Omega^Z,t\in\mathcal{T},h\in\mathcal{H}\notag
\end{align}\label{epsilon}
\end{subequations}
Note that small disturbance terms $\boldsymbol{\varepsilon}$ can not only be regressed from statistical errors in estimation, but also adjusted according to the operators' risk aversion. So hereafter, we name it as \emph{robustness level}, and its associated interval $[\check{\varepsilon}_{ij,t}(\boldsymbol{y}_{ij}),\hat{\varepsilon}_{ij,t}(\boldsymbol{y}_{ij})]$ is \emph{error-related interval}. $\varepsilon^0_{ij}(v)$ in \eqref{epsilon} is the robustness level when line $(i,j)$ is not hardened, and $\varepsilon^h_{ij}(v)$ is defined in Table \ref{Measures}. 
The \emph{Interval B} depicted in Fig. \ref{4Curves} illustrates this interval under  $y_{ij}^{II}=1$.

\begin{figure}
  \centering
  \includegraphics[width=10cm,height=10cm]{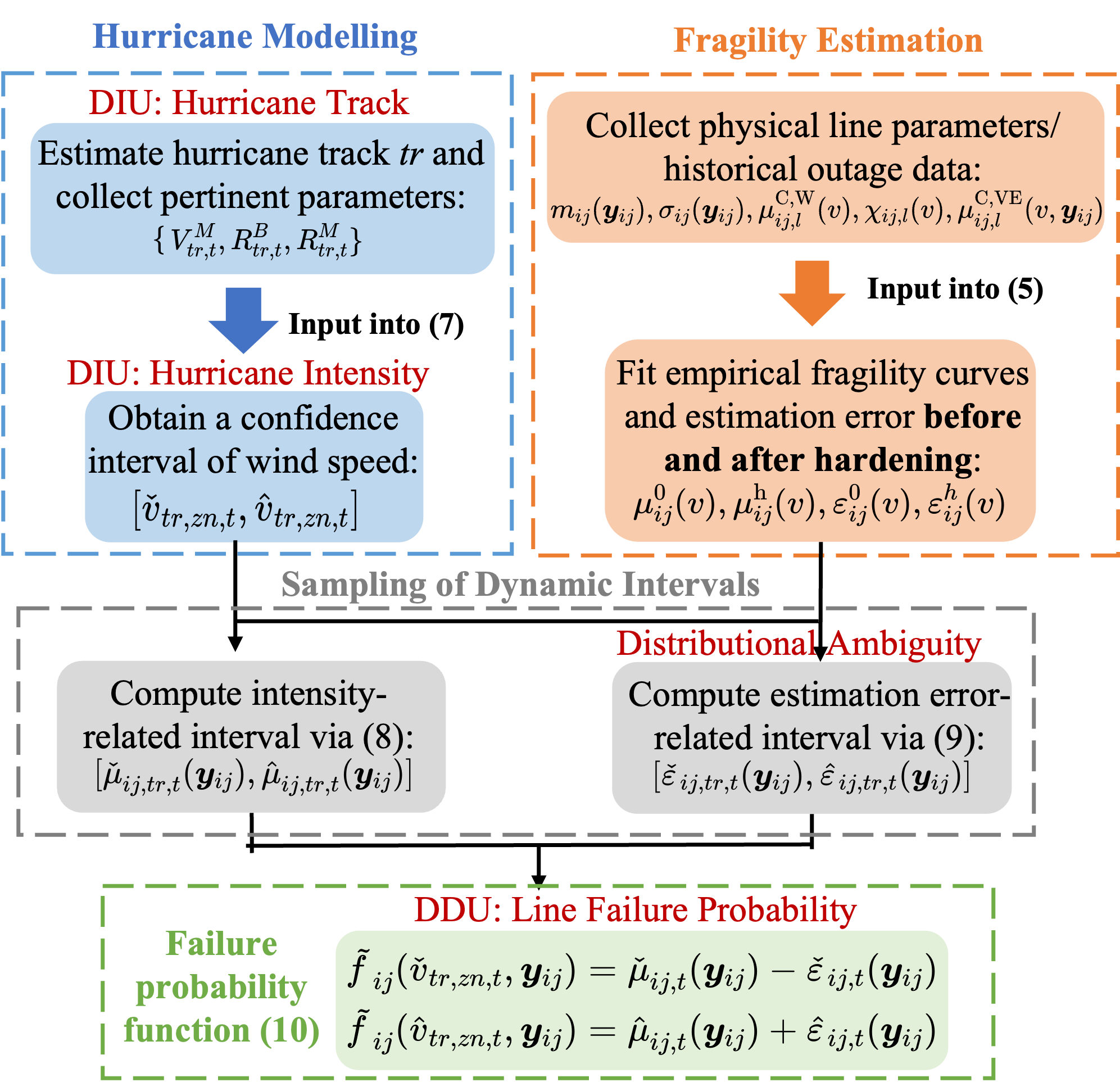}
  \caption{Procedure for parameter derivation of SWDD-ASs}
  \label{procedure}
  \vspace{-2.5ex}
\end{figure}

\subsubsection{Piecewise Linear Failure Probability Function}
Via the establishment of two intervals \eqref{upper/lower probability} and \eqref{epsilon}, the upper and lower values of ambiguous failure probability function in \eqref{ambiguity set-wind speed}-\eqref{supporting set} are derived:\\
\vspace{-5ex}
\begin{subequations}
\begin{align}
    \tilde{f}_{ij}(\check{v}_{tr,zn,t},\boldsymbol{y}_{ij}) = \check{\mu}_{ij,t}(\boldsymbol{y}_{ij}) - \check{\varepsilon}_{ij,t}(\boldsymbol{y}_{ij})
    \\
    \tilde{f}_{ij}(\hat{v}_{tr,zn,t},\boldsymbol{y}_{ij}) = \hat{\mu}_{ij,t}(\boldsymbol{y}_{ij}) + \hat{\varepsilon}_{ij,t}(\boldsymbol{y}_{ij})
\end{align}\label{reformulated DDDAS}
\end{subequations}
\indent The \emph{Interval C} in Figure \ref{4Curves} illustrates the possible range for failure probability of line $(i,j)$ when $y_{ij}^{\rm II}=1$. 
Consequently, through \eqref{fragility-pole}-\eqref{reformulated DDDAS}, constraints in SWDD-ASs \eqref{ambiguity set-wind speed}-\eqref{supporting set} are tractably modeled as piecewise linear functions in $\boldsymbol{y}$ which enables the subsequent reformulations. For a clearer exposition, the step-by-step procedure for constructing DDD-ASs is presented as Figure \ref{procedure}.

\begin{figure}[t!] 
  \centering
  \includegraphics[width=9cm,height=9.3cm]{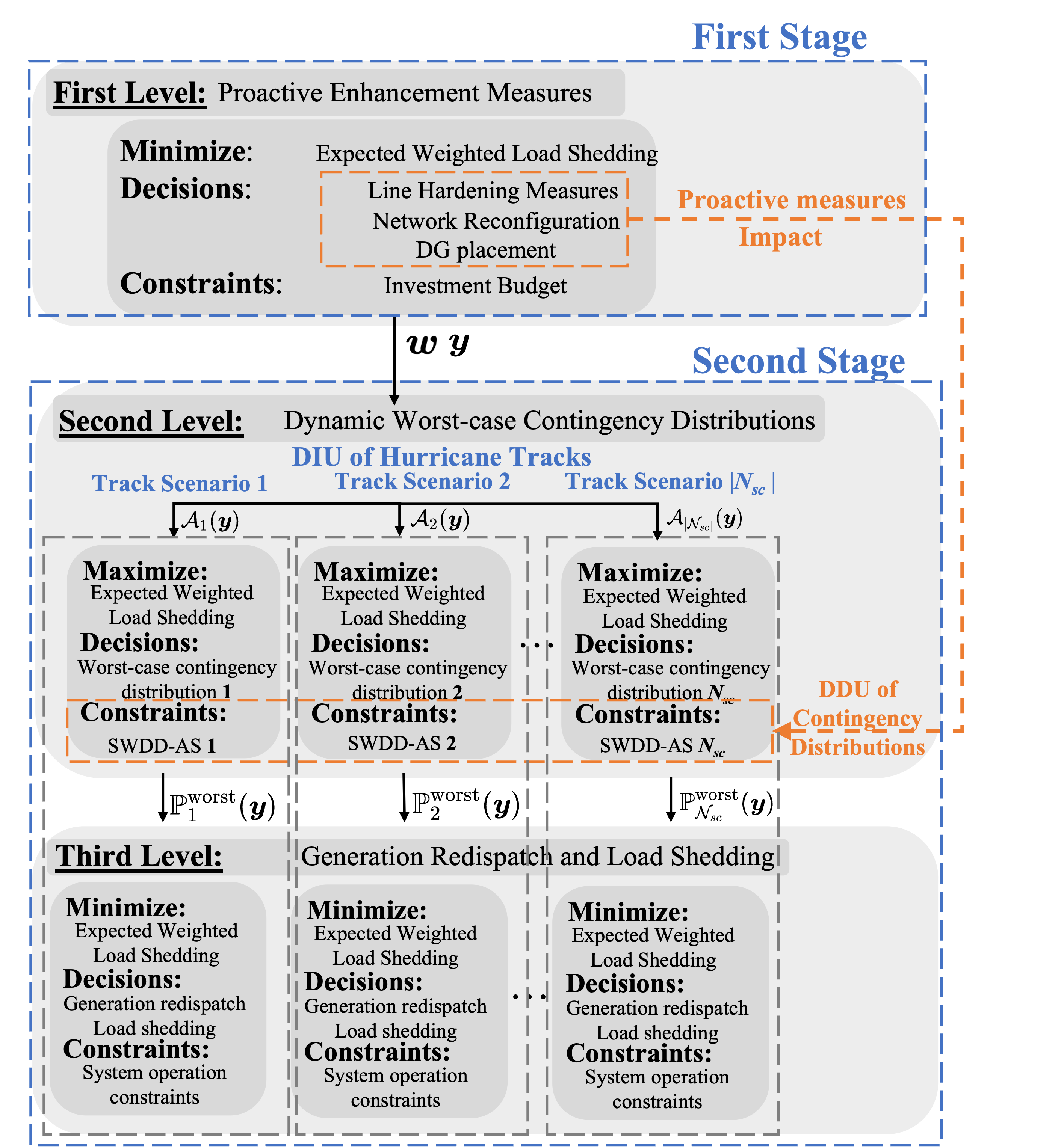}
  \caption{The proposed Two-stage tri-level framework for DD-DRRE model}
  \label{trilevel framework}
  \vspace{-2.5ex}
\end{figure}
\section{Decision-dependent Distributionally Robust Resilience Enhancement Model}
\label{Section 3}

\subsection{A Two-stage Tri-level Framework}

The DD-DRRE model is formulated in a two-stage trilevel framework, as shown in Fig. \ref{trilevel framework}. Specifically, 
the first level constitutes the first stage, aiming to search for the optimal proactive hardening measures before the uncertain contingencies materialize. For each track scenario $tr\in\mathcal{N}_{tr}$ in the second stage, a decision-dependent DRO-based bilevel “max-min” program is nested. In the $tr$-th second level, worst-case contingency distribution $\mathbb{P}^{\rm worst}_{tr}(\boldsymbol{y})$ is identified in the $tr$-th SWDD-AS $\mathcal{A}_{tr}(\boldsymbol{y})$. Subsequently, recourse decisions, namely the generation re-dispatch and load shedding, are derived in the $tr$-th third level based on $\mathbb{P}^{\rm worst}_{tr}(\boldsymbol{y})$.
The detailed formulations are presented as below.

\subsection{Objective function}

Since maintaining the uninterrupted power supply is the most vital task for a resilient distribution system during EWEs \citep{lei2018routing}, our DD-DRRE model aims to find the most effective proactive measures to minimize the EWLS: 
\begin{subequations}
\begin{align}
    {\mathop{\rm min}_{\boldsymbol{w,y}}}\ 
    \!\!\sum_{tr\in\mathcal{N}_{sc}}\!\!\theta_{tr}\cdot
    \Big\{{\mathop{\rm sup}_{\mathbb{P}_{tr}\in
    \mathcal{A}_{tr}(\boldsymbol{y})}}
    {\rm E}_{\mathbb{P}_{tr}}[\varphi_{tr}(\boldsymbol{w},\boldsymbol{y}, \boldsymbol{u}_{tr})]\Big\}
    \label{objective-1st}
\end{align}
where \begin{align}
    \varphi_{tr}(\boldsymbol{w}, \boldsymbol{u}_{tr})=
    {\mathop{\rm min}_{\boldsymbol{z}_{tr}\in \mathcal{Z}_{tr}}}
    \sum_{i\in\Omega^{\rm N}}\sum_{t\in\mathcal{T}}\gamma_i \cdot P^D_{i,t}\cdot \sigma_{tr,i,t}\label{objective-2nd}
\end{align}\label{objective function}
\end{subequations}
In \eqref{objective-1st}, $\boldsymbol{w}$ collects all first-level decision variables except hardening decision $\boldsymbol{y}$.
Vectors $\boldsymbol{u}_{tr}$ and $\boldsymbol{z}_{tr}$ encompass second-level line outage states and third-level decision variables under track $tr$, respectively. 
$\mathcal{Z}_{tr}$ represents the feasible set for $\boldsymbol{z}_{tr}$. 

\subsection{First Level: Line Hardening, DG allocation and Preventive Network Reconfiguration}
The outputs of the first level include hardening decisions for distribution lines, allocation decisions for DERs, as well as preventive network reconfiguration:
{\begin{small}
\begin{align}
    \sum_{h\in\mathcal{H}}\sum_{(i,j)\in\Omega^L}C^h_{ij}y^h_{ij} 
    \le B^{\rm L}
    \label{budget-line}
    \\
    \sum_{i\in\Omega^N}C^G x^G_i \le B^{\rm G}
    \label{budget-dg}
    \\
    \sum_{h\in\mathcal{H}}y^h_{ij}\le 1\qquad\forall (i,j)\in\Omega^{\rm L}
    \label{hardening decision}
    \\
    1 - x^S_{ij}\le s_{ij}\qquad \forall (i,j)\in\Omega^{\rm SW}\label{RCS}
\end{align}
\vspace{-5ex}
\begin{subequations}
\begin{equation}
    \qquad \kappa_i \ge x^{\rm sub}_i \qquad \forall i \in \Omega^N
    \label{root-sub}
\end{equation}
\vspace{-4ex}
\begin{equation}
    \qquad \kappa_i \le x^G_i + x^{\rm sub}_i \qquad \forall i\in\Omega^N
    \label{root}
\end{equation}
\vspace{-4ex}
\begin{equation}
    1\!\!-\!\!M_1\kappa_i\!
    \le\!\!\!\sum_{j|(i,j)\in\Omega^{\rm L}}\!\!\!\!f_{ji}-\!\!\!\!\!\!\!\sum_{j|(i,j)\in\Omega^{\rm L}}\!\!\!\!f_{ij}\le 
    1\!+\!M_1\kappa_i,\forall i\in\Omega^{\rm N}\label{commodity}
\end{equation}
\vspace{-3ex}
\begin{equation}
-M_1 s_{ij}\le f_{ij}
    \!\le\! M_1 s_{ij},\forall (i\!,j)\!\in\! \Omega^{\rm L}\label{on-off}
\end{equation}
\vspace{-4ex}
\begin{equation}
    \sum_{(i,j)\in\Omega^{\rm L}}s_{ij}=|\Omega^{\rm N}|-
    \sum_{i\in\Omega^N}\kappa_{i}
    ,\forall (i,j)\in\Omega^L\label{node-line}
\end{equation}\label{pre-event radiality}
\end{subequations}
\end{small}
}
\!\!\!\!where \eqref{budget-line} and \eqref{budget-dg} are budget constraints on line hardening and DER installation. Constraint \eqref{hardening decision} imposes that only one hardening measure can be implemented on a line $(i,j)$. 
Constraints \eqref{RCS} ensure that only lines with installed switches can be switched off. 
To attain the connectivity and radiality, single commodity flow \eqref{pre-event radiality} is utilized \citep{8080265}. Constraints \eqref{root-sub} and \eqref{root} impose that nodes with substations should be root nodes, and only the nodes installed with DERs or substation can be chosen as root nodes. Constraints \eqref{commodity} ensure the connectivity through virtual commodity balance at each node.  
Constraints \eqref{on-off} state that only switched-on lines can deliver virtual commodity. Constraints \eqref{node-line} together with \eqref{root}-\eqref{on-off} ensure the radiality.

\subsection{Second level: Worst-case contingency distributions within SWDD-ASs}

The second level is to identify the worst-case contingency distribution $\mathbb{P}^{\rm worst}_{tr}(\boldsymbol{y})$ within each SWDD-AS  \eqref{ambiguity set-wind speed}-\eqref{supporting set}. 
The procedure for its construction has been stated in Section \ref{Section 2}. 
\subsection{Third level: Generation Re-dispatch and Load Shedding}
In the third level, DERs are re-dispatched to minimize the EWLS under each $\mathbb{P}^{\rm worst}_{tr}(\boldsymbol{y})$ identified in the second level, while fulfilling the operational constraints:
\begin{subequations}
\begin{align}
(1\!\!-\!\!\sigma^D_{tr,i,t}) P^{\rm D}_{i,t} \!+\! \!\!\!\! \!\sum_{j|(i,j)\in \Omega^{\rm L}}\!\!\! \!\!\!p^{\rm L}_{tr,ij,t}\! -\! \!\!\!\! \!\!\! \sum_{j|(j,i)\in \Omega^{\rm L}}\!\!\!\!p^{\rm L}_{tr,ji,t}
\!&= \!p^{\rm G}_{tr,i,t}+p^{\rm sub}_{tr,i,t} 
\label{powerbalance-p}\\
(1\!\!-\!\!\sigma^D_{tr,i,t}) Q^{\rm D}_{i,t} \!+\!\!\! \!\! \!\sum_{j|(i,j)\in \Omega^{\rm L}}\!\!\!\!\! \!q^{\rm L}_{tr,ij,t}\! -\!\!\! \!\! \!\!\! \sum_{j|(j,i)\in \Omega^{\rm L}}\!\!\!\!q^{\rm L}_{tr,ji,t}
\!&= \!q^{\rm G}_{tr,i,t}+q^{\rm sub}_{tr,i,t}  \!
\label{powerbalance-q}
\end{align}
\label{powerbalance}
\end{subequations}
\vspace{-6.5ex}
\begin{subequations}
\begin{align}
    U^{\rm sqr}_{tr,i,t}-U^{\rm sqr}_{tr,j,t} \leq 2(R_{ij} p^{\rm L}_{tr,ij,t}+&X_{ij} q^{\rm L}_{tr,ij,t}) + 
    M_2 (2-u_{tr,ij,t}-s_{tr,ij,t})
    \\
    U^{\rm sqr}_{tr,i,t}-U^{\rm sqr}_{tr,j,t} \ge 2(R_{ij} p^{\rm L}_{tr,ij,t}+&X_{ij} q^{\rm L}_{tr,ij,t}) + 
    M_2 ( u_{tr,ij,t}+s_{tr,ij,t}-2)
\end{align}
\label{voltagedrop}
\end{subequations}
\vspace{-8.5ex}
\begin{align}
      -s_{ij} \hat P_{ij}^{\rm L} \!\le\! p^{L}_{tr,ij,t} \!\le\! s_{ij} \hat P_{ij}^{\rm L},\ 
    -s_{ij} \hat Q_{ij}^{\rm L} \!\le\! q^{L}_{tr,ij,t} \!\le\! s_{ij} \hat Q_{ij}^{\rm L}\label{switched-line}
\end{align}
\vspace{-8.5ex}
\begin{align}
      -u_{tr,ij,t} \hat P_{ij}^{\rm L} \le p^{L}_{tr,ij,t} \le  & u_{tr,ij,t} \hat P_{ij}^{\rm L},
     -u_{tr,ij,t} \hat Q_{ij}^{\rm L} \le q^{L}_{tr,ij,t} \le u_{tr,ij,t} \hat Q_{ij}^{\rm L} \label{outage-line} \\
     &\qquad
     \forall (i,j)\in \Omega^{\rm L}, t\in \mathcal{T},tr\in\mathcal{N}_{sc}\notag
\end{align}
\vspace{-8.5ex}
\begin{align}
    \check{U}_i^{sqr} &\le U^{\rm sqr}_{tr,i,t}\le \hat U_i^{sqr}\label{voltage}
    \\
    0&\le\sigma_{tr,i,t}\le 1\label{sigma}
\end{align}
\vspace{-8.5ex}
\begin{align}
    0\le p^{\rm G}_{tr,i,t}\le x_{i}\hat P^{\rm G}_{i},
    0\le q^{\rm G}_{tr,i,t}\le x_{i}\hat Q^{\rm G}_{i} 
    \label{DGoutput}
\end{align} 
\vspace{-8.5ex}
\begin{align}
    p^{\rm G}_{tr,i,t}-p^{\rm G}_{tr,i,t-1}\le RU_i,
    p^{\rm G}_{tr,i,t-1}-p^{\rm G}_{tr,i,t}\le RD_i
    \label{DGramping}
    \\
    \forall i\in \Omega^{\rm N},t\in\mathcal{T},tr\in\mathcal{N}_{sc}\notag
\end{align}
\indent DistFlow formulation has been extensively adopted to compute line flows for radial system since it was first developed in \cite{Baran1989Network}. Here the linearized form \eqref{powerbalance}-\eqref{switched-line} is adopted by neglecting the much smaller line losses. Constraints \eqref{powerbalance} are active and reactive power balance equations. Constraints \eqref{voltagedrop} present the voltage drop through the line, where $M_2$ is a sufficiently large number. 
Constraints \eqref{switched-line} ensure that power can only flow through closed lines, and impose the thermal limits. Constraints \eqref{outage-line} impose that power cannot flow through failed lines. 
Voltage limits are stated in \eqref{voltage}. 
Constraints \eqref{sigma} state that the load shedding cannot exceed the predicted demand.
The output limits and inter-temporal ramping rates of DGs are ensured by \eqref{DGoutput} and \eqref{DGramping}, respectively. 
Note that here only the controllable distributed generators (DGs) are considered, but other forms of DERs can be conveniently added to the model.

Therefore, the final DD-DRRE model is defined through objective function \eqref{objective function} with constraints \eqref{ambiguity set-wind speed}-\eqref{supporting set}, \eqref{budget-line}-\eqref{DGramping}, which takes a two-stage trilevel structure that is computationally prohibitive.

\section{Reformulation and Solution Algorithm}
\label{Section 4}
\subsection{Compact Form}
In this section, we introduce the reformulation and solution algorithm for the DD-DRRE model. To simplify the exposition, we first write down the compact form of the DD-DRRE model:
\vspace{-2ex}
\begin{subequations}
\begin{align}
    &{\mathop{\rm min}_{\boldsymbol{w},\boldsymbol{y}}}\ 
     \sum_{tr\in\mathcal{N}_{sc}}\theta_{tr}\cdot
     {\mathop{\rm sup}_{\mathbb{P}_{tr}\in\mathcal{A}_{tr}(\boldsymbol{y})}}\ 
   {\rm E}_{\mathbb{P}_{tr}}[\varphi_{tr}(\boldsymbol{w},\boldsymbol{y}, \boldsymbol{u}_{tr})]\\
    &{\rm{s.t.}}\qquad\qquad \boldsymbol{B}\boldsymbol{w}+\boldsymbol{D}\boldsymbol{y}\le \boldsymbol{b} \label{first-stage constraint}
\end{align}
\label{first-level}
\end{subequations}
\!\!\!\! where 
$\boldsymbol B, \boldsymbol D$ and $\boldsymbol b$ are the coefficient matrices and the right-hand side parameter vector for the first-level constraints, i.e., \eqref{budget-line}-\eqref{pre-event radiality}.
The compact form of $tr$-th SWDD-AS $\mathcal{A}_{tr}(\boldsymbol{y})$ and supporting set $\mathcal{U}_{tr}$ are written as follows:
\begin{align}
    \mathcal{A}_{tr}(\boldsymbol{y})\!=\!&\big\{ \mathbb{P}\in \mathcal{P}(\mathcal{U}_{tr}\big)\big|
    \underline{\boldsymbol{\eta}_{tr}}\!+\!\underline{\boldsymbol{K}_{tr}}\boldsymbol{y}\!\le\! E_{\mathbb{P}_{tr}}[\boldsymbol{1}\!-\!\boldsymbol{u}_{tr}]\!\le\! \overline{\boldsymbol{\eta}_{tr}}\!+\!\overline{\boldsymbol{K}_{tr}}\boldsymbol{y}
    \big\}\label{compact-ambiguity}\\
    \mathcal{U}_{tr}=&\big\{ \boldsymbol{G}\boldsymbol{u}_{tr}\le \boldsymbol{e}_{tr}\big\}\label{compact-supporting}
\end{align}
\!\!\! where $\underline{\boldsymbol{K}_{tr}},\overline{\boldsymbol{K}_{tr}}$ are coefficient matrices, and $\underline{\boldsymbol{\eta}_{tr}},\overline{\boldsymbol{\eta}_{tr}}$ are parameter vectors in $tr$-th SWDD-AS. $\boldsymbol{G}$ and $\boldsymbol{e}_{tr}$ are coefficient matrix and right-hand side vector in $tr$-th supporting set. The \eqref{compact-ambiguity} and \eqref{compact-supporting} correspond to constraints \eqref{ambiguity set-wind speed} and \eqref{supporting set}, respectively. 

The $tr$-th third-level problem with the objective of $\varphi_{tr}(\boldsymbol{w},\boldsymbol{y},\boldsymbol{u}_{tr})$ is presented as below:
\begin{subequations}
\begin{align}
    &\varphi_{tr}(\boldsymbol{w},\boldsymbol{y},\boldsymbol{u}_{tr}) = \mathop{{\rm min}}_{\boldsymbol{z}_{tr}} \boldsymbol{h}^T\boldsymbol{z}_{tr}\\
    &{\rm{s.t.}}\qquad \boldsymbol{H}\boldsymbol{z}_{tr}\le \boldsymbol{l}-\boldsymbol{Jw}-\boldsymbol{Qy}-\boldsymbol{Ru}_{tr}\label{second-stage constraint}
\end{align}
\label{third-level}
\end{subequations}
\!\!\!\! where 
$\boldsymbol{h}$ is the coefficient vector for the objective. $\boldsymbol{l}$ is the right-hand side parameter vector, and $\boldsymbol{H}$, $\boldsymbol{J}$, $\boldsymbol{Q}$, and $\boldsymbol{R}$ are the coefficient matrices in the third-level constraints \eqref{powerbalance}-\eqref{DGramping}.
\subsection{Reformulation of the Scenario-Wise Second Stage}
Unlike fixed ambiguity sets in common DRO problems that only involve DIU, the SWDD-ASs can be altered by decision $\boldsymbol{y}$. Therefore, 
we first reformulate the second stage based on the below proposition by exploring its strong duality property:
\begin{proposition}
For a given first-level decision $(\boldsymbol{w}, \boldsymbol{y})$, the $tr$-th DRO problem in the second stage, i.e., ${\mathop{\rm sup}_{\mathbb{P}_{tr}\in\mathcal{A}_{tr}(\boldsymbol{y})}}
    \!{\rm E}_{\mathbb{P}_{tr}}[\varphi_{tr}(\!\boldsymbol{w}, \boldsymbol{y}, \boldsymbol{u}_{tr}\!)]$\! can be\! reformulated as:
\begin{subequations}
\begin{align}
    &{\mathop{\rm sup}_{\mathbb{P}_{tr}\in\mathcal{A}_{tr}(\boldsymbol{y})}}
    \!\!\!\!\!{\rm E}_{\mathbb{P}_{tr}}[\varphi_{tr}(\boldsymbol{w}, \boldsymbol{y}, \boldsymbol{u}_{tr})]=\!\!\!\!\!\mathop{{\rm min}}_{\boldsymbol{\alpha}_{tr},\boldsymbol{\beta}_{tr}
    \ge 0}\mathop{{\rm max}}_{\boldsymbol{u}_{tr}\in\mathcal{U}_{tr}}\big\{\varphi_{tr}(\boldsymbol{w}, \boldsymbol{y}, \boldsymbol{u}_{tr})+\notag \\&\boldsymbol{\alpha}_{tr}^{\rm T}(\overline{\boldsymbol{\eta}_{tr}}\!\!+\!\!\overline{\boldsymbol{K}_{tr}}\boldsymbol{y}\!\!+\!\!\boldsymbol{u}_{tr}\!-\!\boldsymbol{1})
    \!-\!\boldsymbol{\beta}_{tr}^{\rm T}(\underline{\boldsymbol{\eta}_{tr}}\!+\!\underline{\boldsymbol{K}_{tr}}\boldsymbol{y}\!+\!\boldsymbol{u}_{tr}\!-\!\boldsymbol{1})\big\}\tag{29}
\end{align}
\label{29}
\end{subequations}
\end{proposition}
\vspace{-7ex}
\noindent where $\boldsymbol{\alpha}_{tr}$, $\boldsymbol{\beta}_{tr}$ are dual variables for constraints in\! $tr$-th SWDD-AS. The proof is given in \ref{EC1}. Therefore, by combining \eqref{29}
with the first level, we obtain the following
equivalent form for \eqref{first-level}-\eqref{third-level}:

\vspace{-6ex}
\begin{subequations}
\begin{align}
    \mathop{{\rm min}}_{\boldsymbol{\alpha_{tr}}, \boldsymbol{\beta}_{tr}
    \atop
    \boldsymbol{w},\boldsymbol{y}
    }
    \!\!\!\sum_{tr\in\mathcal{N}_{sc}}\!\!\!\theta_{tr}\cdot
    &\big\{\boldsymbol{\alpha}_{tr}^{\rm T}(\overline{\boldsymbol{\eta}_{tr}}\!+\!\overline{\boldsymbol{K}_{tr}}\boldsymbol{y}\!-\!\boldsymbol{1})\!-\!
    \boldsymbol{\beta}_{tr}^{\rm T}(\underline{\boldsymbol{\eta}_{tr}}\!+\!\underline{\boldsymbol{K}_{tr}}\boldsymbol{y}\!-\!\boldsymbol{1})\}\notag
    \\
    +\!\!\sum_{tr\in\mathcal{N}_{sc}}\!\!\ \theta_{tr}\cdot
    &\big\{\mathop{{\rm max}}_{\boldsymbol{u}_{tr}}\mathop{{\rm min}}_{\boldsymbol{z}_{tr}}\boldsymbol{h}^{\rm T}\boldsymbol{z}_{tr}+(\boldsymbol{\alpha}_{tr}\!-\!\boldsymbol{\beta}_{tr})^{\rm T}\boldsymbol{u}_{tr}\big\}
    \\
    {\rm{s.t.}}\qquad
    &\eqref{first-stage constraint},\eqref{compact-supporting},\eqref{second-stage constraint}
    \\
    &\boldsymbol{\alpha}_{tr}\ge 0,\ \boldsymbol{\beta}_{tr}\ge0
\end{align}\label{reformulation-1}
\end{subequations}
Notably, the reformulation \eqref{reformulation-1} becomes a conventional two-stage robust model with scenario-wise nonlinear second stage. 

\subsection{Customized C\&CG Algorithm-based Solving Procedure}

Due to the fast convergence speed, the C\&CG algorithm is widely adopted to solve the two-stage RO problems \citep{zeng2013solving}. To accommodate the scenario-wise second stage of \eqref{reformulation-1}, the original “min-max-min” problem is decomposed into a “min” master problem (MP) and multiple “max-min” sub-problems (SubPs) that can be computed in a parallel manner. Then MP and mutilple SubPs are computed iteratively until the lower bound identified in MP and upper bound identified in SubPs converge. However, the bilinear terms in both problems that are induced by DDU should be first tackled before the execution. Below presents the procedure for deriving the exact linear reformulations for MP and SubPs. First, for the $n$-th iteration, the MP is formulated as follows:
\begin{subequations}
\begin{align}
    \text{(MP)} \quad 
    &F_{\rm MP}^{(n)*}=\!\!\!\mathop{{\rm min}}_{\boldsymbol{w},\boldsymbol{y},\boldsymbol{\alpha}_{tr}, 
    \atop
    \boldsymbol{\beta}_{tr},
   \boldsymbol{z}_{tr}^{(k)},\gamma}
   \sum_{tr\in\mathcal{N}_{sc}}\!\!\!\theta_{tr}\cdot\Big\{ \boldsymbol{\alpha}_{tr}^{\rm T}(\overline{\boldsymbol{\eta}_{tr}}\!+\!\overline{\boldsymbol{K}_{tr}}\boldsymbol{y}\!-\!\boldsymbol{1})\!-\!\boldsymbol{\beta}_{tr}^{\rm T}(\underline{\boldsymbol{\eta}_{tr}}\!+\!\underline{\boldsymbol{K}_{tr}}\boldsymbol{y}\!-\!\boldsymbol{1})\Big\}\!+\!\gamma \label{MP-1}\\
    &{\rm s.t.}\qquad \ 
    \boldsymbol{Bw}+\boldsymbol{Dy}\le\boldsymbol{b},\boldsymbol{\alpha}_{tr}\ge 0, \boldsymbol{\beta}_{tr}\ge0\label{MP-2}\\
    &\qquad\quad \boldsymbol{Jw}+\boldsymbol{Qy}+\boldsymbol{H}\boldsymbol{z}^{(k)}_{tr}\le \boldsymbol{l}-\boldsymbol{Ru}_{tr}^{(k)}
    \qquad\quad\forall \boldsymbol{u}_{tr}^{(k)}\in \mathcal{S}_{tr},k=1,..,n
    \label{MP-3}
    \\
    &\qquad\quad \gamma \!\ge\! \sum_{tr\in\mathcal{N}_{sc}}\!\!\!\theta_{tr}\cdot\Big\{\boldsymbol{h}^{\rm T}\boldsymbol{z}_{tr}^{(k)}\!\!+\!\!
    (\boldsymbol{\alpha}_{tr}-\boldsymbol{\beta}_{tr})^{\rm T}\boldsymbol{u}^{(k)}_{tr}\Big\}
    \label{MP-4}
\end{align}
\label{MP}
\end{subequations}
\vspace{-1ex}
\!\!\!where $\mathcal{S}_{tr}$ is the worst-case scenario set whose components $\boldsymbol{u}_{tr}^{(k)}(k=1,..,n)$ are identified in $tr$-th SubP through $n$-times iteration. $\gamma$ and $\boldsymbol{z}_{tr}^{(k)}(k=1,..,n)$ are auxiliary decision variables. 
McCormick envelopes is utilized here to linearize bilinear terms introduced by DDU, i.e., $\boldsymbol{\alpha}_{tr}^{\rm T}\overline{\boldsymbol{K}_{tr}}\boldsymbol{y}$ and $\boldsymbol{\beta}_{tr}^{\rm T}\underline{\boldsymbol{K}_{tr}}\boldsymbol{y}$, and the linearized MP is denoted as MP'. Related details are presented in \ref{EC2}.

\indent After computing linearized MP' in the $n$-th iteration, we can obtain their $n$-th solutions $\boldsymbol{w}^{(n)*}$, $\boldsymbol{y}^{(n)*}$, $\boldsymbol{\alpha}_{tr}^{(n)*}$, and $\boldsymbol{\beta}_{tr}^{(n)*}$ and feed them into the $tr$-th subproblem ($tr$-th SubP) as below:
\begin{subequations}
\begin{align}
    \text{($tr$-th SubP)}\ 
   & F_{{\rm SP},tr}^{(n)*} \!\!= \!\!\mathop{{\rm max}}_{\boldsymbol{u}_{tr}}\mathop{{\rm min}}_{\boldsymbol{z}_{tr}}\Big\{\boldsymbol{h}^{\rm T}\!\boldsymbol{z}_{tr}\!\!+\!\!
    (\boldsymbol{\alpha}_{tr}^{(n)*}\!\!\!-\!\boldsymbol{\beta}_{tr}^{(n)*})^{\rm T}\boldsymbol{u}_{tr}\Big\}\label{SP-1}\\
    {\rm s.t.}\qquad
    &\boldsymbol{G}\boldsymbol{u}_{tr}\le\boldsymbol{e}_{tr}\label{SP-2}\\
    &\boldsymbol{H}\boldsymbol{z}_{tr}\le \boldsymbol{l}-\boldsymbol{Jw}^{(n)*}-\boldsymbol{Qy}^{(n)*}-\boldsymbol{Ru}_{tr}\label{SP-3}
\end{align}
\label{SP}
\end{subequations}
As the third-level “min” problem is convex with respect to $\boldsymbol{z}_{tr}$, \eqref{SP} can be equivalently transformed into its dual problem based on the strong duality condition:
\begin{subequations}
\begin{align}
    \text{($tr$-th SubP')}
    \quad
    F_{{\rm SP},tr}^{(n)*} =
    \mathop{{\rm max}}_{\boldsymbol{u}_{tr},
    \atop
    \boldsymbol{\pi}_{tr},
    \boldsymbol{\tau}_{tr}}
    \Big\{(\boldsymbol{\alpha}_{tr}^{(n)*}-
    &\boldsymbol{\beta}_{tr}^{(n)*})^{\rm T}\boldsymbol{u}_{tr}+
    (\boldsymbol{l}\!-\!\boldsymbol{Jw}^{(n)*}-\boldsymbol{Qy}^{(n)*})^{\rm T}\boldsymbol{\pi}_{tr}\!-\!\boldsymbol{1}^{\rm T}\boldsymbol{\tau}_{tr}\Big\}
    \\
    {\rm s.t.}\qquad &\boldsymbol{G}\boldsymbol{u}_{tr}\le\boldsymbol{e}_{tr}
    \\
    &\boldsymbol{H}^{\rm T}\boldsymbol{\pi}_{tr}=\boldsymbol{h},\boldsymbol{\pi}\le 0
    \\
    -{M}_3\boldsymbol{u}_{tr}\le\boldsymbol{\tau}_{tr}\le 0,&
    -{M}_3(\boldsymbol{1}-\boldsymbol{u}_{tr})\le \boldsymbol{\tau}_{tr}-\boldsymbol{R}^{\rm T}\boldsymbol{\pi}_{tr}\le 0
    \label{bigM2}
\end{align}
\label{SP'}
\end{subequations}
\!\!where $\boldsymbol{\pi}_{tr}$ is the dual variable for constraints \eqref{SP-3}. Constraints \eqref{bigM2} are used to linearize the bilinear terms $\boldsymbol{u}_{tr}^T\boldsymbol{R}^{\rm T}\boldsymbol{\pi}_{tr}$ via big-M method, where $\boldsymbol{\tau}_{tr}$ is auxiliary variable and $M_3$ is a sufficiently large number.

Since the MP' and $tr$-th SubP' are both exactly formulated as MILPs, 
they can be directly solved by commercial solvers. Here we customize the traditional C\&CG algorithm to solve our problem with multiple scenario-wise SubPs. The pseudo-code is presented in Algorithm \ref{algorithm}. Notably, since the load shedding is considered, feasible solution must exist in each $tr$-th SubP'. Thus, the upper and lower bounds will finally convergence to a singleton within finite iterations. 
\begin{algorithm}
\caption{Customized C\&CG Algorithm}\label{algorithm}
{\bf{Step 1.}}\quad {\bf{Initialization.}} Set lower bound $\rm LB\leftarrow -\infty$, upper bound $\rm UB\leftarrow +\infty$, iteration time $k\leftarrow 0$, set of contingency $\mathcal{S}_{tr}=\emptyset$ and optimality gap $\delta=0.01\%$\;

{\bf{Step 2.}}\quad
\While{$|\frac{\rm UB-LB}{\rm UB}|>\delta$}{
    Update $k=k+1$
\\Solve (MP') to obtain the optimal solutions, $\boldsymbol{w}^{(k)*},$ $\boldsymbol{y}^{(k)*},$ $\boldsymbol{\alpha}_{tr}^{(k)*}, \boldsymbol{\beta}_{tr}^{(k)*},\gamma^{(k)*}$, and \!optimal\! value\! $F^{(k)*}_{\rm MP}$
\\
    Update ${\rm LB} \leftarrow F^{(k)*}_{\rm MP}$
    \\
    \For{$tr=1$ \KwTo$|\mathcal{N}_{sc}|$}
    {Solve $tr$-th SubP' to obtain the optimal solution, $\boldsymbol{u}_{tr}^{(k)*}$,
    and optimal value $F^{(k)*}_{{\rm SP},tr}$
    \\
    Update $\mathcal{S}_{tr}=\mathcal{S}_{tr}\cup{\boldsymbol{u}_{tr}^{(k)*}}$
    }
    Update ${\rm{UB}} =$ ${\rm{min}}\{{\rm{UB}},F^{(k)*}_{\rm MP}-\gamma^{(k)*}+\sum_{tr\in\mathcal{N}_{sc}}\theta_{tr}F^{(k)*}_{{\rm SP},tr}\}$}
    {\bf{Step 3.\quad Terminate.}}
    Return $\rm UB$,$\boldsymbol{w}^{(k)*}$, and $\boldsymbol{y}^{(k)*}$
\end{algorithm}

\section{Derivation of the Worst-case Distributions and Evaluation Methods}
\label{Section 5}

In this section, to facilitate the comparisons with two conventional counterparts, i.e., SP- and RO-based models, we derive closed forms of worst-case contingency distributions in SWDD-ASs, as well as two useful metrics showing hidden risks of the two counterparts.

\subsection{Worst-case Distribution of Contingencies}

Stress tests are critical to evaluate the performances of strategies under uncertain environment. To conduct effective testing, it is necessary to
derive the extremal distributions within the SWDD-ASs that achieve the worst-case expectation.
Here the worst-case distributions are derived according to the below proposition \citep{8998220} (the proof is given in \ref{EC3}):
\begin{proposition}
Suppose that the customized C {\normalfont\&}CG algorithm
terminates at the $N$-th iteration with optimal solutions
$(\boldsymbol{w}^{(N)*},\boldsymbol{y}^{(N)*},\boldsymbol{\alpha}_{tr}^{(N)*}, \boldsymbol{\beta}_{tr}^{(N)*},\{\boldsymbol{z}_{tr}^{(k)*}\}_{k=1,..,N},\gamma^{(N)*})$. Then, if we resolve MP with variables $\boldsymbol{w}, \boldsymbol{y}$ and $\boldsymbol{z}^{k}_{tr}$ fixed at $\boldsymbol{w}^{(N)*}, \boldsymbol{y}^{(N)*}$ and $\boldsymbol{z}_{tr}^{(k)*}$,
respectively, the dual optimal solutions associated
with constraints \eqref{MP-4}, denoted as ${\Xi_{tr}^k}$ $(k=1,...,N)$, characterize
the worst-case contingency probability distribution under hurricane track $tr$. Mathematically, we have:\\
\end{proposition}
\vspace{-7ex}
\begin{align}
    \mathbb{P}_{tr}^{\rm worst}(\boldsymbol{u}_{tr}=\boldsymbol{u}_{tr}^{(k)}|&\boldsymbol{w}^{(N)*},\boldsymbol{y}^{(N)*},\{\boldsymbol{z}_{tr}^{(k)*}\}_{k=1,..,N})= {\Xi_{tr}^k} \quad \forall tr\in\mathcal{N}_{tr}, k=1,..,N
\end{align}

\subsection{The Value of Distributional Ambiguity and The Value of Moment Information} 
\begin{table}[]\scriptsize
\centering
\caption{Comparison of Different Modeling Framework}
\label{Measures}
\vspace{1ex}
\begin{tabular}{cccc}
\hline\hline
\begin{tabular}[c]{@{}c@{}}\bf{Modeling} \\ \bf{Framework}\end{tabular} & \begin{tabular}[c]{@{}c@{}}\bf{Contingency} \\ \bf{Quantification}\end{tabular}           & \begin{tabular}[c]{@{}c@{}}\bf{Optimality}\\ \bf{Criterion}\end{tabular}                        & \begin{tabular}[c]{@{}c@{}}\bf{Optimal}\\  \bf{Value}\end{tabular}                   \\ \hline
\bf{DD-DRRE }                                                      & $\mathcal{A}_{tr}(\boldsymbol{y})$          & \begin{tabular}[c]{@{}c@{}}Worst Distribution\\  ($\mathbb{P}^{\rm worst}_{tr}$)\end{tabular} & $F^{\rm dro}_{\boldsymbol{\varepsilon}}(\boldsymbol{w}^{\rm dro}\!,\!\boldsymbol{y}^{\rm dro})$ \\
\bf{DD-SRE  }                                                      & $\mathbb{P}^{\rm em}_{tr}(\boldsymbol{y})$ &\begin{tabular}[c]{@{}c@{}}Multiple Scenarios\\  ($\boldsymbol{u}^{\rm sp}_{tr,1},..,\boldsymbol{u}^{\rm sp}_{tr,S} $)\end{tabular}                 & $F^{\rm sp}(\boldsymbol{w}^{\rm sp}\!,\!\boldsymbol{y}^{\rm sp})$   \\
\bf{RRE    }                                                       & $\mathcal{U}_{tr}$                                                                     & \begin{tabular}[c]{@{}c@{}}Worst Scenario\\  ($\boldsymbol{u}^{\rm worst}_{tr}$)\end{tabular}                                                             & $F^{\rm ro}(\boldsymbol{w}^{\rm ro}\!,\!\boldsymbol{y}^{\rm ro})$   \\\hline \hline
\end{tabular}
\end{table}
The out-of-sample disappointment quantifies the probability that the actual EWLS of the candidate decision under the unknown true distribution exceeds its expectation. 
Here, two metrics are proposed to quantitatively exhibit this likely disappointment of conventional SP and RO-based strategies. Two counterpart strategies are compared, namely the decision-dependent stochastic resilience enhancement (DD-SRE) model (based on SP) and robust resilience enhancement (RRE) model (based on RO). Their prior assumptions and optimality criterion are listed in the Table \ref{Measures}. Notably, the DD-SRE accounts for the DDU of empirical contingency distributions $\mathbb{P}^{\rm em}_{tr}$ but neglects its ambiguity, while RRE neglects the distributions' information and their associated decision-dependency by only finding the worst scenario in the uncertainty set $\mathcal{U}_{tr}$. $F_{\boldsymbol{\varepsilon}}^{\rm dro}(\boldsymbol{w},\boldsymbol{y}),F^{\rm sp}(\boldsymbol{w},\boldsymbol{y}),F^{\rm ro}(\boldsymbol{w},\boldsymbol{y})$ are defined as the optimal value functions under three modeling framework with first-level decision fixed at $(\boldsymbol{w},\boldsymbol{y})$, while $(\boldsymbol{w}^{\rm dro},\boldsymbol{y}^{dro}),(\boldsymbol{w}^{\rm sp},\boldsymbol{y}^{sp})$, and $(\boldsymbol{w}^{\rm ro},\boldsymbol{y}^{ro})$ are optimal fist-level solutions of the three models. Therefore, the two metrics 
are defined as below:
\begin{equation}
\begin{aligned}
    \text{VoDA}_{\boldsymbol{\varepsilon}}=\frac{F_{\boldsymbol{\varepsilon}}^{\rm dro}(\boldsymbol{w}^{\rm sp})-
    F_{\boldsymbol{\varepsilon}}^{\rm dro}(\boldsymbol{w}^{\rm dro})}
    {F_{\boldsymbol{\varepsilon}}^{\rm dro}(\boldsymbol{w}^{\rm dro})}
    \\ 
    \text{VoMI}_{\boldsymbol{\varepsilon}}=\frac{F_{\boldsymbol{\varepsilon}}^{\rm dro}(\boldsymbol{w}^{\rm ro})-F_{\boldsymbol{\varepsilon}}^{\rm dro}(\boldsymbol{w}^{\rm dro})}{F_{\boldsymbol{\varepsilon}}^{\rm dro}(\boldsymbol{w}^{\rm dro})}
    \label{Evaluation}
\end{aligned}    
\end{equation} 

The $\text{VoDA}_{\boldsymbol{\varepsilon}}$ in \eqref{Evaluation} defines the \emph{value of distributional ambiguity (VoDA)}, i.e., the maximal extra value planners might pay by believing that the true contingency distribution is $\mathbb{P}^{\rm em}_{tr}$, when $\mathbb{P}_{tr}^{\rm worst}$ accurately represents the contingency distributions (measured in percentage to facilitate the comparisons). Similarly, the second equation in \eqref{Evaluation} defines the \emph{value of moment information (VoMI)}, i.e., the maximal extra value planner might pay by being overly conservative when SWDD-ASs accurately consider the ambiguity of contingency distributions. 
Notably, VoDA and VoMI are both parametrized by robustness level $\boldsymbol{\varepsilon}$, which controls the sizes of SWDD-ASs. 
Therefore, they can help system planners understand how valuable or risky the SP- and RO-based solutions are in the DRO setting and explicitly show their hidden risks in different external conditions. 
\section{Case Studies}
\label{Section 6}
\subsection{IEEE 33-Node Test System and Parameter Setting}
This section presents the numerical tests of the proposed method. Two test systems are investigated: modified IEEE 33-node and 123-node test systems. We first present the parameters and results of the 33-node system. 
Due to the page limit, some results of IEEE 33-node system are eliminated from the main manuscript and the full results are presented in \ref{EC4}. 


For the hurricane modeling, four hurricane track scenarios with different probabilities 
are considered. 
The hurricane duration is 24h with time resolution of 1h. 
According to the method stated in \cite{javanbakht2014risk}, 100 sets of hurricane parameters $\{V^M_{tr,t},R^B_{tr,t},R^M_{tr,t}\}$ are randomly generated and fitted for each track to get the wind speed confidence interval $[\check{v}_{tr,zn,t},\hat{v}_{tr,zn,t}]$. 
The network hardening budget $B^{\rm H}$ for IEEE 33-node system is set to \$$5.86\times10^5$. 
Moreover, the cost for upgrading distribution poles is \$6000/pole \cite{fang2019adaptive}, and the span of two consecutive poles is 50 meters. The vegetation management is \$12500/km 
\cite{8329529}. The active and reactive capacities of DGs are all set to be 300kW and 175kVar, and there are 4 DGs for allocation. 

\begin{figure}
  \centering
  \includegraphics[width=12cm,height=4.6cm]{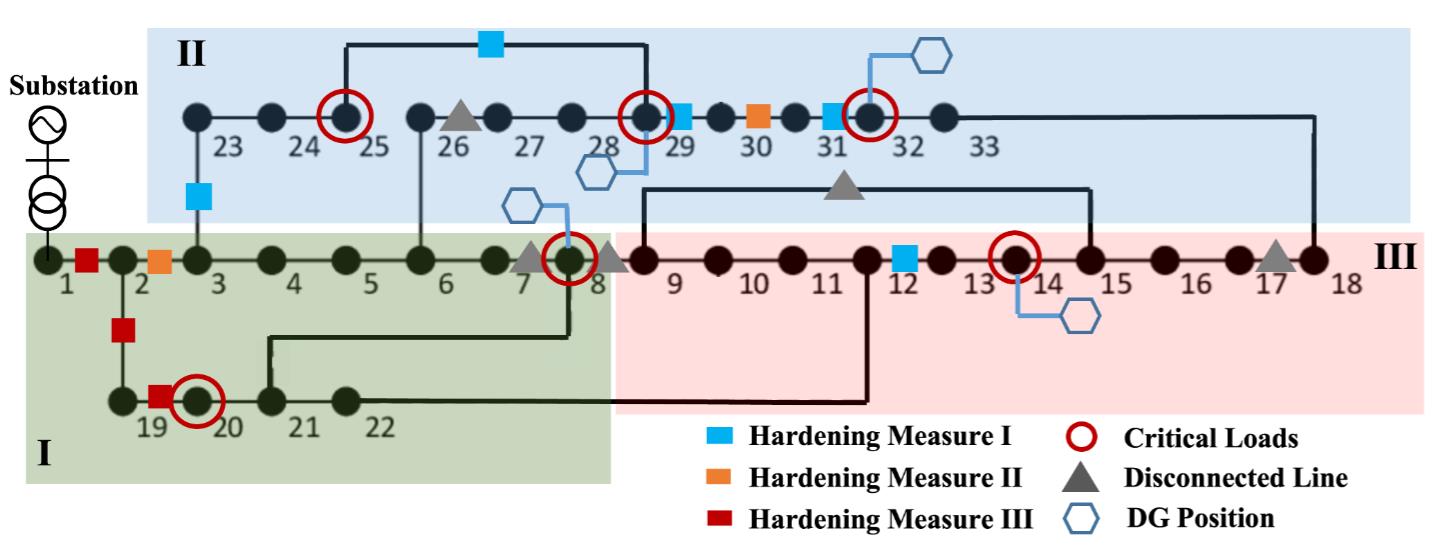}\\
  \caption{Optimal enhancement strategy for modified IEEE 33-node system derived from the DD-DRRE model}
  \label{33-configuration}
\end{figure}

Fig. \ref{33-configuration} shows the modified IEEE 33-node test system, which is divided into 3 zones based on their geographical footprint. The system originally contains 1 substation at node 1 and 37 distribution lines. 
The system voltage base is 12.66 kV, and voltage lower and upper limits are set as 0.9 p.u. and 1.1p.u.. 
Node 8, 14, 20, 25, 29 and 31 are chosen as critical load nodes. To reflect different aging states of distribution lines, the empirical fragility curves of different lines before and after hardening or vegetation management are randomly tuned based on data from \cite{1046899} and \cite{hughes2021damage}.
Robust level $\boldsymbol{\varepsilon}$ in the nominal test is set as $0.3$ times of the empirical failure rate. 
The active and reactive loads at each node are randomly generated from intervals $[60,420]$kW and $[20,200]$kVars. 
The priority weights of non-critical load nodes are randomly generated within the interval $[1,5]$. However, the weights of critical load nodes that have higher requirements for power supply continuity are set as 50. 

The computation is performed in Matlab 2021a and solved by Gurobi 9.1 via CVX toolbox, on an Intel Core i5-6500 CPU with 16 GB RAM PC. The convergence tolerance is set as 0.01\%.
\subsection{Results of the Proposed DD-DRRE Model}

\indent  The enhancement results derived from the proposed DD-DRRE model is illustrated in Fig. \ref{33-configuration}, with an EWLS of 20220.6 kW. Due to the limited DGs' capabilities to provide uninterrupted power to critical loads on a local level, the lines from the substation to critical load node 20 are hardened using the most effective measure III, while vegetation management is implemented on lines 3-23 and 25-29 to increase the survivability of critical load node 25. 
Notably, the lines between node 28 and 31 are hardened to increase the probability of power supply to node 25. Additionally, the hardening measures also benefit the power supply of noncritical loads such as node 12, which connects two downstream branches.




\vspace{-2.5ex}
\begin{table}[]\scriptsize
\centering
\caption{Different prior assumptions of four strategies}
\label{4-Strategies}
\vspace{1ex}
\begin{tabular}{lccc}
\hline \hline
      \textbf{Strategy} & \begin{tabular}[c]{@{}c@{}}\textbf{Distributional} \\ \textbf{Ambiguity}\end{tabular} & \begin{tabular}[c]{@{}c@{}}\textbf{Decision-dependency} \\ \textbf{of Distributions}\end{tabular} & \begin{tabular}[c]{@{}c@{}}\textbf{Uncertainty of} \\\textbf{Hurricane Intensity} \end{tabular} \\ \hline
\textbf{1} (DD-DRRE) &\small \checkmark               &\small  \checkmark              &\small  \checkmark                       \\
\textbf{2} (DD-SRE) &{\tiny \XSolid}     &\small  \checkmark             &\small  \checkmark                       \\
\textbf{3} (RRE) &{\tiny \XSolid}                   &{\tiny \XSolid}             &\small  \checkmark                       \\
\textbf{4} (DD-DRRE with FHI) &\small  \checkmark                   &\small \checkmark             &{\tiny \XSolid}                      \\ \hline \hline
\end{tabular}
\end{table}

\subsection{In-Sample and Out-of-Sample Evaluations}
To demonstrate the potential value of incorporating the DDU and distributional ambiguity, we examine four strategies derived from four counterpart models with distinct prior assumptions. According to TABLE \ref{4-Strategies}, Strategy 1 is generated from our proposed model. Strategy 2 is derived from the SP-based DD-SRE model, 
and Strategy 3 is drawn from the RO-based RRE model. 
By setting the wind speed to its expected value, Strategy 4 eliminates the random property of hurricane intensity, which is denoted as DD-DRRE with fixed hurricane intensity (FHI). Due to the page limit, the detailed enhancement results of Strategy 2 and 3 are presented in \ref{EC4} with further analyses.  

TABLE \ref{33-Comparison} summarizes their in-sample and out-of-sample performances. The second column (OBJ) presents the in-sample EWLS. 
In the presence of DDU and distributional ambiguity, we test their out-of-sample EWLS using two test sets: 1) WCD set, whose scenarios are generated using the worst-case distribution obtained in proposition 1; and 2) RGD set, whose scenarios are obtained from random distributions within SWDD-ASs. The third and fourth columns of TABLE IV respectively report the weighted averages under these two tests.

\begin{table}\footnotesize
\centering
\caption{Comparison of EWLS of four strategies in IEEE 33-node system}
\label{33-Comparison}
\vspace{1ex}
\begin{tabular}{cccc}
\hline\hline
\textbf{Strategy} & \textbf{OBJ} (kW) & \textbf{WCD} (kW) & \textbf{RGD} (kW) \\ \hline
\textbf{1}        & 20220.6                                                     & 19494.3                                                     & 18290.7                                                     \\
\textbf{2}        & 18094.7                                                    & 22314.8                                                     & 20167.9                                                     \\
\textbf{3}        & 40515.4                                                     & 21791.5                                                     & 22578.6                                                     \\
\textbf{4}                 & 17046.6                                                     & 32714.6                                                     & 29942.4                                                     \\ 
\hline\hline
\end{tabular}
\end{table}

\begin{figure}
\centering
\includegraphics[width=5cm,height=5cm]{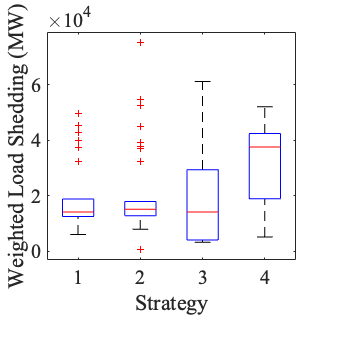}
\caption{Boxplot of WLS under WCD in IEEE 33-node system}
\label{33-boxplot}
\end{figure}

As shown in TABLE \ref{33-Comparison}, although Strategy 1 obtained by our method does not have the lowest in-sample expectation, it does have the lowest EWLS under WCD and RGD tests.
The lower in-sample EWLS in Strategy 2 than in Strategy 1 and 3 is mostly due to its optimism in the accuracy of empirical distributions, but this also results in a $14.5\%$ and a $23.4\%$ increase in results under WCD and RGD tests.
For strategy 3, the in-sample EWLS is the highest owing to the pessimistic prior assumption that the worst-case contingency scenario will occur.  
Out-of-sample performances of Strategy 4 are the worst, demonstrating the critical role of capturing the random nature of hurricane intensity.

Moreover, in order to compare the performances of four strategies under worst-case distributions, Fig. \ref{33-boxplot} depicts a boxplot graph of the weighted load shedding (WLS) under different scenarios in WCD set. With a comparable narrow span and positive skewness, our proposed Strategy 1 can produce relatively consistent and stable results with few outliers. While Strategy 2 has a similar pattern to Strategy 1, it has a higher median and more scattered outliers, indicating that it is more vulnerable to contingency misspecification.  
With regards to Strategy 3, its boxplot graph has a much larger span than those of Strategies 1 and 2, indicating that the distribution-free RO-based strategy with no DDU consideration cannot guarantee a stable result. 
Finally, the wide span and negative skewness of boxplot for Stratey 4 demonstrate the necessity of more accurate hurricane knowledge.

\begin{figure}

\centering

\subfigure[VoDA (\%)]{
\includegraphics[width=6cm,height=4.5cm]{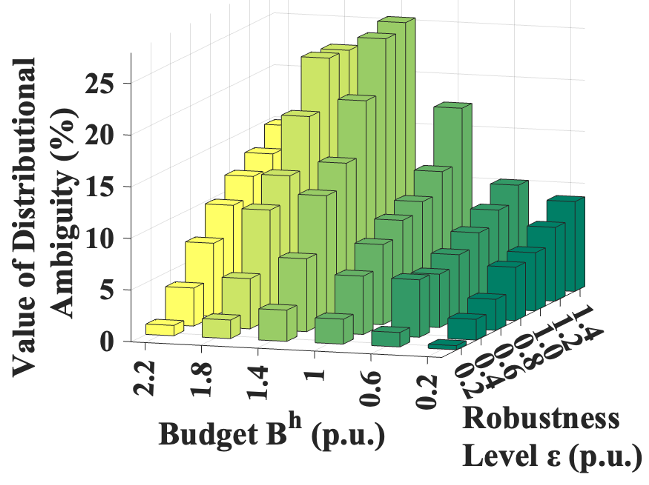}}
\subfigure[VoMI (\%)]{
\includegraphics[width=6cm,height=4.5cm]{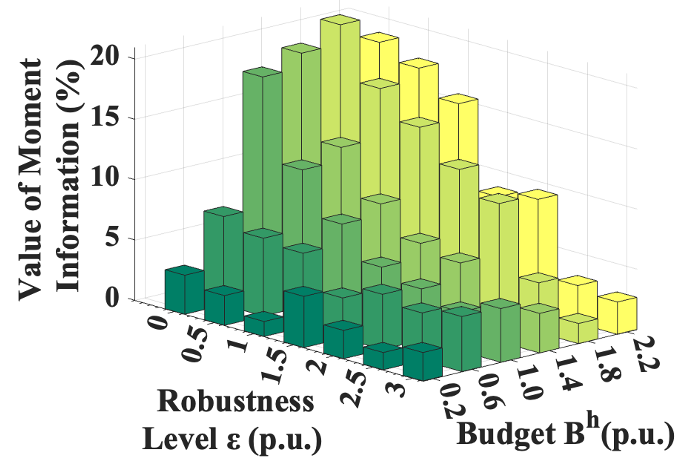}}
\caption{Expected VoDA and VoMI under various  budgets and robustness levels in IEEE 33-node system}
\label{POPP}
\end{figure}
\subsection{Results of VoDA and VoMI}
In the presence of ambiguity and DDU, the hidden risks and applicability of the prevalent ambiguity-free SP- and DDU-free RO-based counterparts are investigated by computing VoDA and VoMI under different line hardening budgets $B^{\rm L}$ and robustness levels $\boldsymbol{\varepsilon}$. The results are depicted in Fig. \ref{POPP}, which have the following properties.

First, given any hardening budget, the VoDA represented in Fig. \ref{POPP}(a) grows monotonically in $\boldsymbol{\varepsilon}$, reaching 26.14\% at $\boldsymbol{\varepsilon} = 1.4\ \rm p.u.$ when the budget is $\$1.4 \times 5.16 \times 10^5$. The rationale is that an increased $\boldsymbol{\varepsilon}$ implies that either the empirical fragility estimation is less trustworthy or the planner is more risk-averse. 
In this regard, simply adopting the SP-based strategy will inevitably result in a higher retrospective regret. 
On the other hand, VoMI in Fig. \ref{POPP}(b) exhibits an overall non-increasing pattern in terms of $\boldsymbol{\varepsilon}$. 
This is because mistrust in empirical contingency distributions and associated DDU will result in an increasing homogeneity between our proposed DRO-based model and the RO-based model.

Second, given a fixed $\boldsymbol{\varepsilon}$, both the VoDA and VoMI first grow with the budget $B^{\rm L}$, but subsequently decline after reaching their peaks around $1.4\ \rm p.u.$ and $1.8\ \rm p.u.$, respectively. 
The declining trends after the peak imply that excessive investment might help smooth the effects of model misspecification.

Finally, we also notice that there are several circumstances making the performances of our proposed strategy relatively indistinguishable from its counterparts. 
When budgets is inadequate, both the VoDA and VoMI are trivial, due to the limited hardened line. 
Besides this, when empirical estimation of component failure rates is highly accurate (i.e., $\boldsymbol{\varepsilon}$ is small), the SP-based DD-SRE model is acceptable. Moreover, the RO-based RRE model is a reasonable substitute when the underlying contingency distributions and associated DDU are highly unpredictable. 
Nevertheless, 
our proposed approach is obviously superior 
in most conditions with moderate budgets and risk-aversion levels.


\subsection{Results on IEEE 123-Node System}

\begin{figure}
  \centering
  \includegraphics[width=11cm,height=10cm]{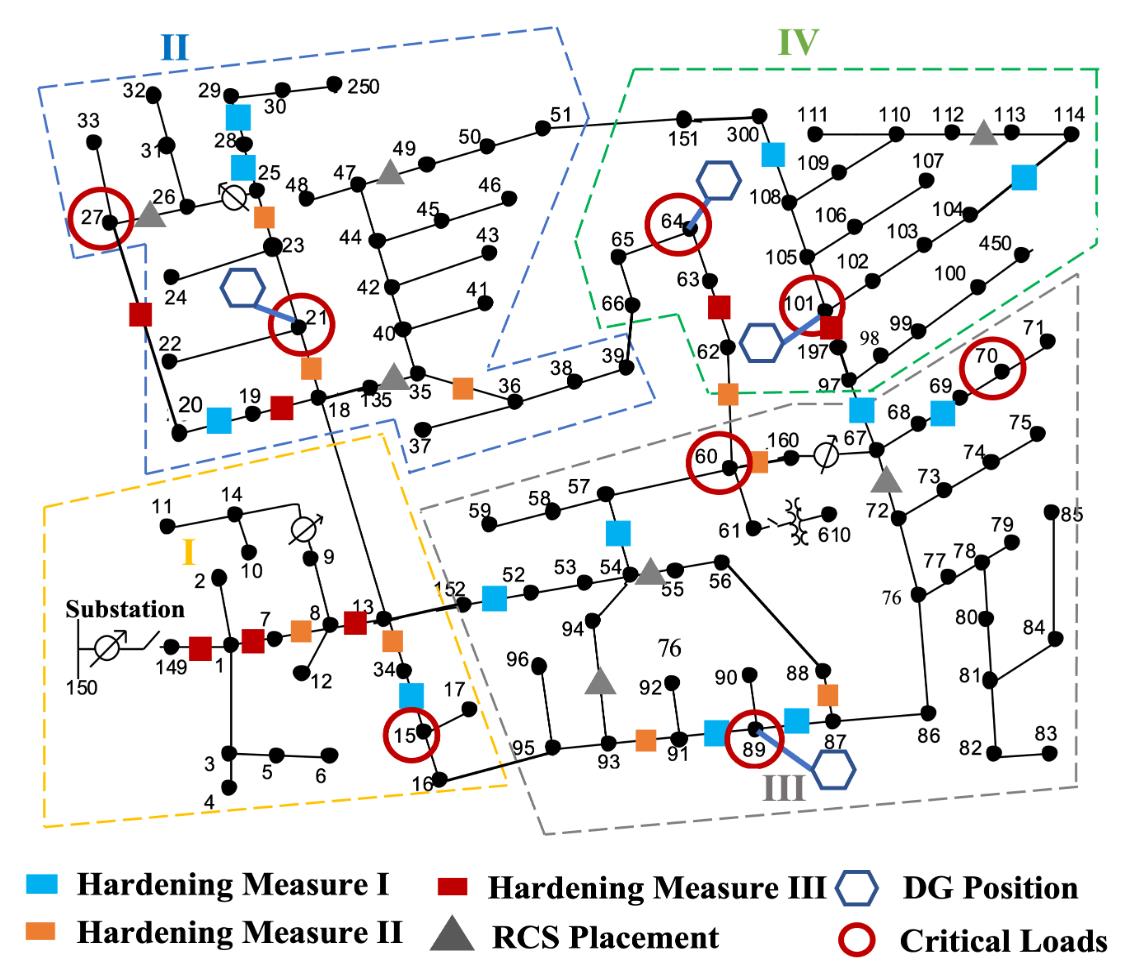}\\
  \caption{Optimal enhancement strategy in modified IEEE 123-node system derived from the DD-DRRE model}
  \label{123-configuration}
\end{figure}

The scalability of the presented formulation is further verified using the modified IEEE 123-node test system. The system is geographically divided into four zones, with the substation located at node 150. The hardening budget $B^{\rm H}$ for 123-node test systems is set to \$$2.16\times10^6$. 
There are also 4 DGs for allocation, with active and reactive capacities of  300kW and 175kVar. 
The active and reactive loads at each node are randomly generated from $[20,180]$kW and $[10,120]$kVars. Apart from the stated settings and network parameters, all other parameters used in this test are identical to those in IEEE 33-node system.

Fig. \ref{123-configuration} presents the enhancement result. Notably, the lines from DERs to critical loads 60 and 70 are not entirely hardened due to the budget constraint, which contributes significantly to the  EWLS. Furthermore, the moderate hardening actions carried out on the lines in less affected zone III illustrate the need of geographical partition.

\begin{table}\small
\caption{Comparison of WLS of four strategies in IEEE 123-node system}
\label{123-comparison}
\vspace{1ex}
\centering
\begin{tabular}{cccc}
\hline\hline
\textbf{Strategy} & \textbf{EXP (kW)} & \textbf{WCD (kW)} & \textbf{RGD (kW)} \\ \hline
\textbf{1}        & 52217.0                                                     & 52238.5                                                     & 50834.7                                                     \\
\textbf{2}        & 45236.7                                                     & 60560.9                                                     & 53969.3                                                     \\
\textbf{3}        & 128324.2                                                    & 57515.4                                                     & 56135.7                                                     \\
\textbf{4}                 & 39153.8                                                     & 93554.2                                                     & 73246.6                                                     \\ \hline\hline
\end{tabular}
\end{table}
\begin{figure}
\centering
\includegraphics[width=5cm,height=5cm]{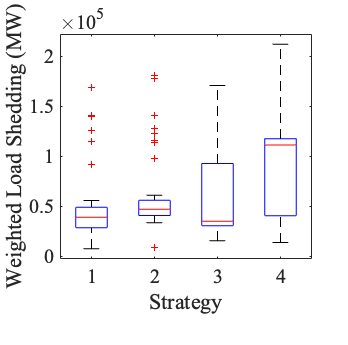}
\caption{Boxplot of WLS under WCD in IEEE 123-node system}
\label{123-boxplot}
\end{figure}

In addition, TABLE \ref{123-comparison} summarizes the in-sample and out-of-sample performances of our proposed strategy and its three counterparts. Our proposed Strategy 1 still performs the best under WCD and RGD tests. 
When the worst-case contingency distributions in SWDD-ASs materialize, Strategy 2 is more unreliable than Strategies 1 and 3, although Strategy 3 is relatively insensitive to distribution perturbation within SWDD-ASs. 
Again, Strategy 4 has the highest EWLS in both the WCD and RGD tests. The boxplot under the WCD test in Fig. \ref{123-boxplot} also demonstrates the relatively steady performance of our strategy.

\begin{figure}
\centering

\subfigure[VoDA (\%)]{
\includegraphics[width=6cm,height=4.5cm]{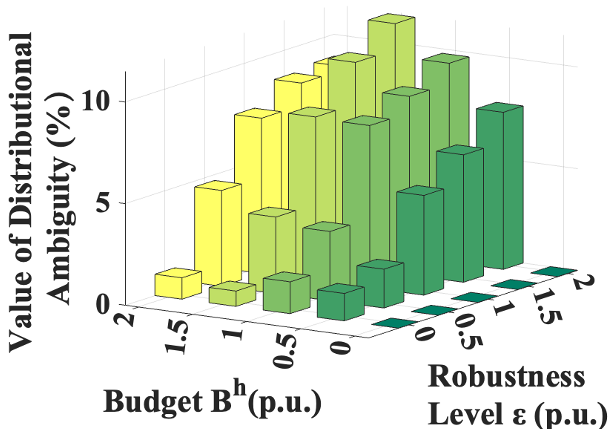}}
\subfigure[VoMI (\%)]{
\includegraphics[width=6cm,height=4.5cm]{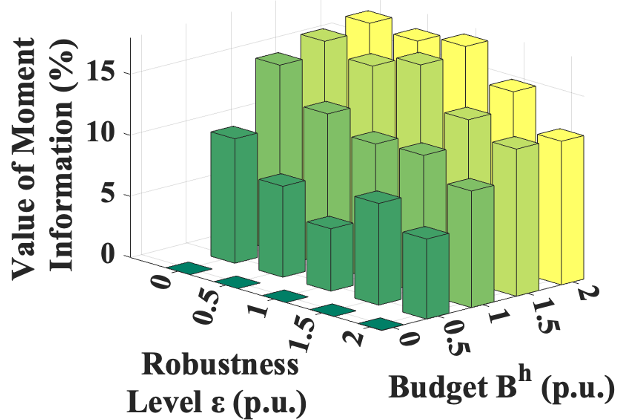}}

\caption{Expected VoDA and VoMI under various hardening budgets and robustness levels in IEEE 123-node test system}
\label{123-POPP}
\vspace{-2ex}
\end{figure}

VoDA and VoMI are computed for 123-node system and Fig. \ref{123-POPP} reports the results. 
With fixed budget, the VoDA rises with $\boldsymbol{\varepsilon}$, while the VoMI presents non-increasing trend on the whole. 
When budgets are increased, the VoDA first grows monotonically but then begins to decline after peaking around $B^{\rm h}=$1.5 p.u., showing the benefits of increasing network investment to hedge against model misspecification. 
In contrast to the 33-node system, the VoMI, on the other hand, tends to increase monotonously in terms of budget. This is likely due to the the increased scalability necessitates a more aggressive hardening budget to show the smoothing effects.

Overall, the results from the 123-node system are consistent with those from the 33-node system, illustrating the scalability and generality of our proposed DD-DRRE model.

\section{Conclusion}
\label{Section 7}
Accounting for the interdependence between hardening decisions and uncertain contingency under EWEs, as well as its associated ambiguous estimation, this paper proposes a novel two-stage trilevel DD-DRRE model to assist distribution grids in developing proactive resilience enhancement strategies. 
By presenting a step-by-step procedure for constructing SWDD-ASs, we are able to capture both the DDU of ambiguous contingency distributions and the DIU of hurricane track and intensity. 
To alleviate the computational burdens imposed by the non-convex and nonlinear formulation, we derive its equivalent form using its strong duality property regarding the distribution and exact linearization techniques. This permits the subsequent execution of a customized C\&CG algorithm. 
From numerical studies, we find that simultaneous quantification of the DDU and distributional ambiguity can greatly safeguard distribution grids against
real-world misspecification in contingency modeling, compared with its prevalent counterparts. 
Furthermore, the computation of VoDA and VoMI quantify the hidden price that planners may extra pay when simply relying on ambiguity-free SP-based and DDU-free RO-based models, 
drawing attention to the great value of our proposed method in most cases.





\bibliography{mybibfile}

\newpage
\appendix 
\setcounter{table}{0}
\setcounter{figure}{0}


\setcounter{page}{1}

\section{Proof of Proposition 1}
\label{EC1}
\noindent {\textbf{Proposition 1.}}
For a given first-level decision $(\boldsymbol{w}, \boldsymbol{y})$, the $tr$-th DRO problem in the second stage, i.e., ${\mathop{\rm sup}_{\mathbb{P}_{tr}\in\mathcal{A}_{tr}(\boldsymbol{y})}}
    \!{\rm E}_{\mathbb{P}_{tr}}[\varphi_{tr}(\!\boldsymbol{w}, \boldsymbol{y}, \boldsymbol{u}_{tr}\!)]$\! can be\! reformulated as:
\begin{subequations}
\begin{align}
    &{\mathop{\rm sup}_{\mathbb{P}_{tr}\in\mathcal{A}_{tr}(\boldsymbol{y})}}
    \!\!\!\!\!{\rm E}_{\mathbb{P}_{tr}}[\varphi_{tr}(\boldsymbol{w}, \boldsymbol{y}, \boldsymbol{u}_{tr})]=\!\!\!\!\!\mathop{{\rm min}}_{\boldsymbol{\alpha}_{tr},\boldsymbol{\beta}_{tr}
    \ge 0}\mathop{{\rm max}}_{\boldsymbol{u}_{tr}\in\mathcal{U}_{tr}}\big\{\varphi_{tr}(\boldsymbol{w}, \boldsymbol{y}, \boldsymbol{u}_{tr})+\notag \\&\boldsymbol{\alpha}_{tr}^{\rm T}(\overline{\boldsymbol{\eta}_{tr}}\!\!+\!\!\overline{\boldsymbol{K}_{tr}}\boldsymbol{y}\!\!+\!\!\boldsymbol{u}_{tr}\!-\!\boldsymbol{1})
    \!-\!\boldsymbol{\beta}_{tr}^{\rm T}(\underline{\boldsymbol{\eta}_{tr}}\!+\!\underline{\boldsymbol{K}_{tr}}\boldsymbol{y}\!+\!\boldsymbol{u}_{tr}\!-\!\boldsymbol{1})\big\}
\end{align}
\end{subequations}

\begin{proof}

First, if $\boldsymbol{w}$ is fixed, the $tr$-th second level can be rewritten as follows:
\begin{subequations}
\begin{align}
    {\mathop{\rm sup}_{\mathbb{P}_{tr}\in\mathcal{A}_{tr}(\boldsymbol{y})}}\!\!\!\!\!
   {\rm E}_{\mathbb{P}_{tr}}[&\varphi_{tr}(\boldsymbol{w},\boldsymbol{y}, \boldsymbol{u}_{tr})]\!=\! \mathop{{\rm max}}_{\mathbb{P}_{tr}} \int_{\mathcal{U}_{tr}}\!\!\!\!\!\varphi_{tr}(\boldsymbol{w},\boldsymbol{y},\boldsymbol{u}_{tr})d\mathbb{P}_{tr}
   \\
    {\rm{s.t.}}\qquad &\int_{\mathcal{U}_{tr}}d\mathbb{P}_{tr}=1
    \\
    &\int_{\mathcal{U}_{tr}}(\boldsymbol{1}-\boldsymbol{u}_{tr})d\mathbb{P}_{tr}\le \overline{\boldsymbol{\eta_{tr}}}+\overline{\boldsymbol{K}_{tr}}\boldsymbol{y}
    \\
    &\int_{\mathcal{U}_{tr}}(\boldsymbol{1}-\boldsymbol{u}_{tr})d\mathbb{P}_{tr}\ge \underline{\boldsymbol{\eta}_{tr}}+\underline{\boldsymbol{K}_{tr}}\boldsymbol{y}
\end{align}
\label{middle-level-extend}
\end{subequations}
\indent We can find that at least one interior solution exists for \eqref{middle-level-extend}, for example the distribution where all lines have $\boldsymbol{u}=\boldsymbol{0}$ with the probability of $\frac{\overline{\boldsymbol{\eta}}+\underline{\boldsymbol{\eta}}}{2}+\frac{\overline{\boldsymbol{K}}+\underline{\boldsymbol{K}}}{2}\boldsymbol{w}$. Thus, the Slater's condition holds. Meanwhile, $\varphi_{tr}(\boldsymbol{w},\boldsymbol{y},\boldsymbol{u}_{tr})$ obviously has an upper bound, so strong duality holds. Therefore, it can be equivalently reformulated as its dual problem:
\begin{subequations}
\begin{align}
    &\mathop{{\rm min}}_{\nu_{tr},\boldsymbol{\alpha}_{tr}\ge 0,
    \atop
    \boldsymbol{\beta}_{tr}\ge 0} \!\!\!
    \nu_{tr} + 
    \boldsymbol{\alpha}_{tr}^{\rm T}(\overline{\boldsymbol{\eta}_{tr}}+\overline{\boldsymbol{K}_{tr}}\boldsymbol{y})-\boldsymbol{\beta_{tr}}^{\rm T}(\underline{\boldsymbol{\eta}_{tr}}+\underline{\boldsymbol{K}_{tr}}\boldsymbol{y})\label{middle-level-obj-re}
    \\
    &{\rm s.t.} \quad
    \nu_{tr} + (\boldsymbol{\alpha}_{tr}-\boldsymbol{\beta}_{tr})^{\rm T}(\boldsymbol{1}-\boldsymbol{u}_{tr})\ge \varphi_{tr}(\boldsymbol{w},\boldsymbol{y},\boldsymbol{u}_{tr}) 
    \forall \boldsymbol{u}_{tr}\in\mathcal{U}_{tr}\label{max-constraint}
\end{align}

The \eqref{max-constraint} can be further transformed into the following:
\begin{align}
    \nu_{tr}
    \!\ge\!\! 
    \mathop{{\rm max}}_{\boldsymbol{u}_{tr}\in\mathcal{U}_{tr}}\!
    \big\{\varphi_{tr}(\boldsymbol{w},\boldsymbol{y},\boldsymbol{u}_{tr}) \!-\! (\boldsymbol{\alpha}_{tr}\!-\!\boldsymbol{\beta}_{tr})^{\rm T}(\boldsymbol{1}\!-\!\boldsymbol{u}_{tr})\big \}\label{re-reformulation}
\end{align}
\end{subequations}
Subsequently, by substituting \eqref{re-reformulation} into  \eqref{middle-level-obj-re}, we get:
\begin{align}
\begin{aligned}
    &{\mathop{\rm sup}_{\mathbb{P}_{tr}\in
    \atop
    \mathcal{A}_{tr}(\boldsymbol{y})}}
    \!\!{\rm E}_{\mathbb{P}}[\varphi_{tr}(\boldsymbol{w},\boldsymbol{y}, \boldsymbol{u}_{tr})]\!\!=\!\!\!\!\mathop{{\rm min}}_{\boldsymbol{\alpha_{tr}}\ge 0,
    \atop
    \boldsymbol{\beta}_{tr}
    \ge 0}\!\!\mathop{{\rm max}}_{\boldsymbol{u}_{tr}\in\mathcal{U}_{tr}}\big\{\varphi_{tr}(\boldsymbol{w},\boldsymbol{y},\boldsymbol{u}_{tr}) 
    \\
    &+\!\boldsymbol{\alpha}_{tr}^{\rm T}\!(\overline{\boldsymbol{\eta}_{tr}}\!+\!\!\overline{\boldsymbol{K}_{tr}}\boldsymbol{y}\!+\!\boldsymbol{u}_{tr}\!\!-\!\!\boldsymbol{1})
    \!\!-
    \!\!\boldsymbol{\beta}_{tr}^{\rm T}\!(\underline{\boldsymbol{\eta}_{tr}}\!\!+\!\underline{\boldsymbol{K}_{tr}}\boldsymbol{y}\!+\!\boldsymbol{u}_{tr}\!\!-\!\!\boldsymbol{1})\!\big\}
\end{aligned}
\end{align}
      
\end{proof}

\section{Linearized Reformulation of MP based on McCormick Envelope}
\label{EC2}
The bilinear terms introduced by DDU, i.e., $\boldsymbol{\alpha}_{tr}^{\rm T}\overline{\boldsymbol{K}_{tr}}\boldsymbol{y}$ and $\boldsymbol{\beta}_{tr}^{\rm T}\underline{\boldsymbol{K}_{tr}}\boldsymbol{y}$ can be reformulated through McCormick Envelope as below:

\vspace{-1.5ex}
\begin{small}
\begin{subequations}
\begin{align}
    \forall r,l:  &(\overline{\boldsymbol{\alpha}_{tr}})_{r}(\overline{\boldsymbol{K}_{tr}})_{rl}y_{l} \le \rho_{1,tr,r,l} \!\!\le\!\! (\underline{\boldsymbol{\alpha}_{tr}})_{r}(\overline{\boldsymbol{K}_{tr}})_{rl}y_{l}\\
    &(\overline{\boldsymbol{\beta}_{tr}})_{r}(\underline{\boldsymbol{K}_{tr}})_{rl}y_{l} \le \rho_{2,tr,r,l} \!\!\le\!\! (\underline{\boldsymbol{\beta}_{tr}})_{r}(\underline{\boldsymbol{K}_{tr}})_{rl}y_{l}\\
    &({\boldsymbol{\alpha}_{tr}})_{r}(\overline{\boldsymbol{K}_{tr}})_{rl}\!-\!(\underline{\boldsymbol{\alpha}_{tr}})_{r}(\overline{\boldsymbol{K}_{tr}})_{rl}(1\!-\!y_l)\!\le\! \rho_{1,tr,r,l} \!\le\!\notag
    \\
    &\qquad\quad({\boldsymbol{\alpha}_{tr}})_{r}(\overline{\boldsymbol{K}_{tr}})_{rl}\!-\!(\overline{\boldsymbol{\alpha}_{tr}})_{r}(\overline{\boldsymbol{K}_{tr}})_{rl}(1\!-\!y_l)
    \\
    &({\boldsymbol{\beta}_{tr}})_{r}(\underline{\boldsymbol{K}_{tr}})_{rl}\!-\!(\underline{\boldsymbol{\beta}_{tr}})_{r}(\underline{\boldsymbol{K}_{tr}})_{rl}(1\!-\!y_l)\!\le\! \rho_{2,tr,r,l} \!\le\!\notag
    \\
    &\qquad\quad({\boldsymbol{\beta}_{tr}})_{r}(\underline{\boldsymbol{K}_{tr}})_{rl}\!-\!(\overline{\boldsymbol{\beta}_{tr}})_{r}(\underline{\boldsymbol{K}_{tr}})_{rl}(1\!-\!y_l)
\end{align}
\label{McCormick}
\end{subequations}
\end{small}
\noindent \!\!\!\!where $(\underline{\boldsymbol{K}_{tr}})_{rl}/(\overline{\boldsymbol{K}_{tr}})_{rl}$ are $(r,l)$-th components of matrix $\underline{\boldsymbol{K}_{tr}}$/$\overline{\boldsymbol{K}_{tr}}$. The $\underline{\boldsymbol{\alpha}_{tr}}/\overline{\boldsymbol{\alpha}_{tr}}$ and $\underline{\boldsymbol{\beta}_{tr}}/\overline{\boldsymbol{\beta}_{tr}}$ are lower/upper bounds for $\boldsymbol{\alpha}_{tr}$ and $\boldsymbol{\beta}_{tr}$, respectively, and the subscript $r$/$l$ denote their $r$-th/$l$-th  components. Note that \eqref{McCormick} are the exact reformulation for \eqref{MP}, as $\boldsymbol{y}$ is binary. Therefore, via auxiliary variables $\boldsymbol{\rho}_{1,tr}=[\rho_{1,tr,r,l},\forall r,l]^{\rm T}$ and $\boldsymbol{\rho}_{2,tr}=[\rho_{2,tr,r,l},\forall r,l]^{\rm T}$, we get the MILP-based master problem (MP') as below:

\vspace{-2ex}
\begin{small}
\begin{subequations}
\begin{align}
    \text{ (MP')}\quad\quad
    &F_{\rm MP}^{(n)*}=
    \!\!\!\!\!\!\mathop{{\rm min}}_{\boldsymbol{w},\boldsymbol{\alpha}_{tr}, \boldsymbol{\beta}_{tr},\boldsymbol{z}^{(k)},
    \atop
    \boldsymbol{\rho}_{1,tr},\boldsymbol{\rho}_{2,tr},\gamma_{tr}}
    \sum_{tr\in\mathcal{N}_{sc}}\!\!\!\theta_{tr}\cdot
    \Big\{\boldsymbol{\alpha}_{tr}^{\rm T}(\overline{\boldsymbol{\eta}_{tr}}-\boldsymbol{1})\notag
    \\
    -\boldsymbol{\beta}_{tr}^{\rm T}&(\underline{\boldsymbol{\eta}_{tr}}-\boldsymbol{1})
    +\boldsymbol{1}^{\rm T}\boldsymbol{\rho}_{1,tr}-\boldsymbol{1}^{\rm T}\boldsymbol{\rho}_{2,tr}\Big\}+\gamma
\end{align}
\vspace{-4ex}
\begin{align}
    {\rm s.t.\qquad \eqref{MP-2},\eqref{MP-3},\eqref{MP-4},\eqref{McCormick}} 
\end{align}
\label{MP'}
\end{subequations}
\end{small}

\section{Proof of Proposition 2}
\label{EC3}
\noindent{\textbf{Proposition 2.}}
Suppose that the customized C\&CG algorithm terminates at the $N$-th iteration with optimal solutions
$(\boldsymbol{w}^{(N)*},\boldsymbol{y}^{(N)*},\boldsymbol{\alpha}_{tr}^{(N)*}, \boldsymbol{\beta}_{tr}^{(N)*}, \{\boldsymbol{z}_{tr}^{(k)*}\}_{k=1,..,N},\gamma^{(N)*})$. Then, if we resolve (MP) with variables $\boldsymbol{w}, \boldsymbol{y}$ and $\boldsymbol{z}_{tr}$ ($k=1,..,N$) fixed at $\boldsymbol{w}^{(N)*}, \boldsymbol{y}^{(N)*}$ and $\boldsymbol{z}_{tr}^{(k)*}$ ($k=1,..,N$),
respectively, the dual optimal solutions associated
with constraints \eqref{MP-4}, denoted as ${\Xi_{tr}^k}$ $(k=1,...,N)$, characterize
the worst-case contingency probability distribution under hurricane track scenario $tr$. Mathematically, we have:\\
\vspace{-4ex}
\begin{align}
    \mathbb{P}_{tr}^{\rm worst}(\boldsymbol{u}_{tr}=\boldsymbol{u}_{tr}^{(k)}|&\boldsymbol{w}^{(N)*},\boldsymbol{y}^{(N)*},\boldsymbol{z}^{(k)*})= {\delta^{(k)}_{tr}} \\&\forall tr\in\mathcal{N}_{tr}, k=1,..,N
    \notag
\end{align}

\begin{proof}

By fixing $\boldsymbol{w}, \boldsymbol{y}$ and $\boldsymbol{z}_{tr}$ ($k=1,..,N$) at $\boldsymbol{w}^{(N)*}, \boldsymbol{y}^{(N)*}$ and $\boldsymbol{z}_{tr}^{(k)*}$ ($k=1,..,N$),
we derive the equivalent dual reformulation of (MP) as below:
\begin{subequations}
\begin{align}
    \mathop{{\rm max}}_{{\delta}_{tr}^{(k)}\ge0}
    &\sum_{tr\in\mathcal{N}_{sc}}\theta_{tr}\cdot\sum_{k=1}^N
    {\delta}^{(k)}_{tr}
    \cdot(\boldsymbol{h}\boldsymbol{z}_{tr}^{(k)})
    \label{WCD-obj}
    \\
    {\rm s.t.}\qquad & 
    \qquad\eqref{MP-2},\eqref{MP-3},
    \\
    \forall{tr}\in\mathcal{N}_{sc}:\quad &\quad\sum_{k=1}^{N}{\delta}_{tr}^{(k)}=1, \label{WCD-1}
    \\
    \sum_{k=1}^{N}{\delta}_{tr}^{(k)}&(1-\boldsymbol{u}_{tr}^{(k)})\le \overline{\boldsymbol{\eta}_{tr}}+\overline{\boldsymbol{K}_{tr}}\boldsymbol{y}^{(N)*}\label{WCD-2}
    \\
     \sum_{k=1}^{N}{\delta}_{tr}^{(k)}&(1-\boldsymbol{u}_{tr}^{(k)})\ge \underline{\boldsymbol{\eta}_{tr}}+\underline{\boldsymbol{K}_{tr}}\boldsymbol{y}^{(N)*}\label{WCD-3}
\end{align}
\end{subequations}
Therefore, \eqref{WCD-1}-\eqref{WCD-3} yield the worst probability distirbuiton supported on scenarios $\{\boldsymbol{u}_{tr}^{(k)}\}_{k=1,..,N}$ with probability $\{\delta^{(k)}_{tr}\}_{k=1,..,N}$ under each track $tr$, and \eqref{WCD-obj} corresponds to the weighted expected objective under these worst distributions.
\end{proof}

\section{Full Results of IEEE 33-Node Test System}
\label{EC4}
\subsection{Results of the Proposed DD-DRRE Model}
\begin{figure}
  \centering
  \includegraphics[width=12cm,height=4.4cm]{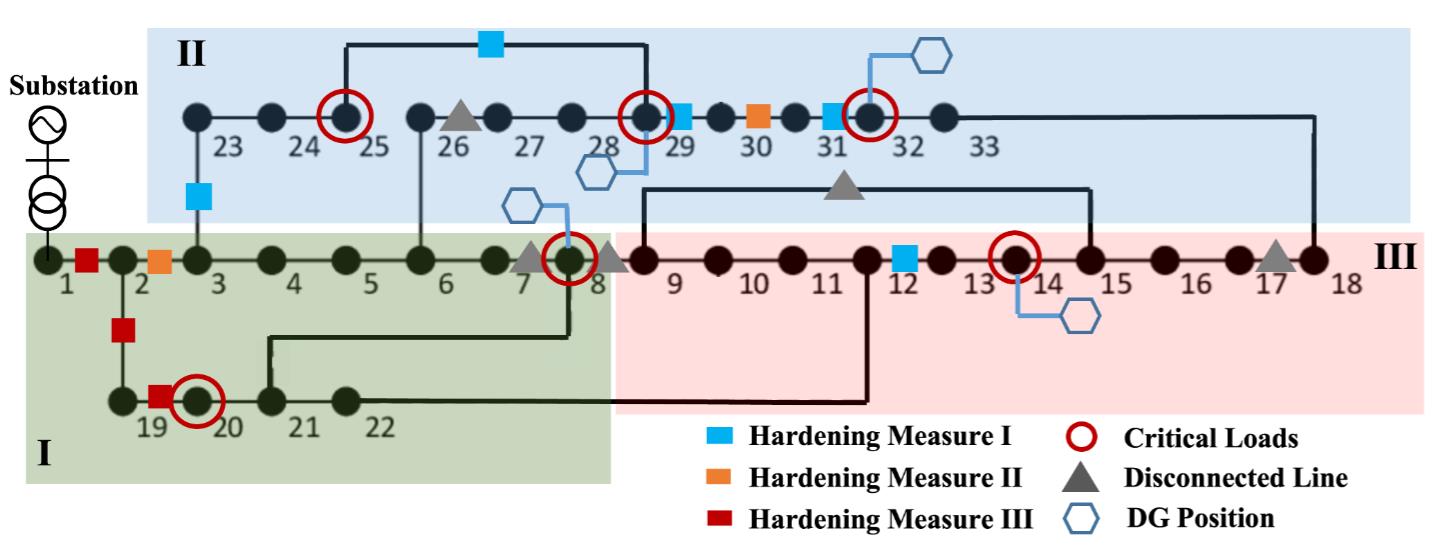}\\
  \caption{Optimal enhancement strategy for modified IEEE 33-node system derived from the DD-DRRE model}
  \label{33-configuration2}
\end{figure}
Fig. \ref{33-configuration2} shows the modified IEEE 33-node test system, which is divided into 3 zones based on their geographical footprint. The system originally contains 1 substation at node 1 and 37 distribution lines. 
The system voltage base is 12.66 kV, and voltage lower and upper limits are set as 0.9 p.u. and 1.1p.u.. 
Node 8, 14, 20, 25, 29 and 31 are chosen as critical load nodes.

\indent  The enhancement results derived from the proposed DD-DRRE model is also illustrated in Fig. \ref{33-configuration2}, with an EWLS of 20220.6 kW. Due to the limited DGs' capabilities to provide uninterrupted power to critical loads on a local level, the lines from the substation to critical load node 20 are hardened using the most effective measure III, while vegetation management is implemented on lines 3-23 and 25-29 to increase the survivability of critical load node 25. 
Notably, the lines between node 28 and 31 are hardened to increase the probability of power supply to node 25. Additionally, the hardening measures also benefit the power supply of noncritical loads such as node 12, which connects two downstream branches.

\vspace{-2ex}
\subsection{Contributions of DG Allocation and Proactive Network Reconfiguration} 
\begin{figure}

\centering

\subfigure[Expected weighted load curve under hurricane track scenario 1]{
\includegraphics[width=6.3cm,height=4.8cm]{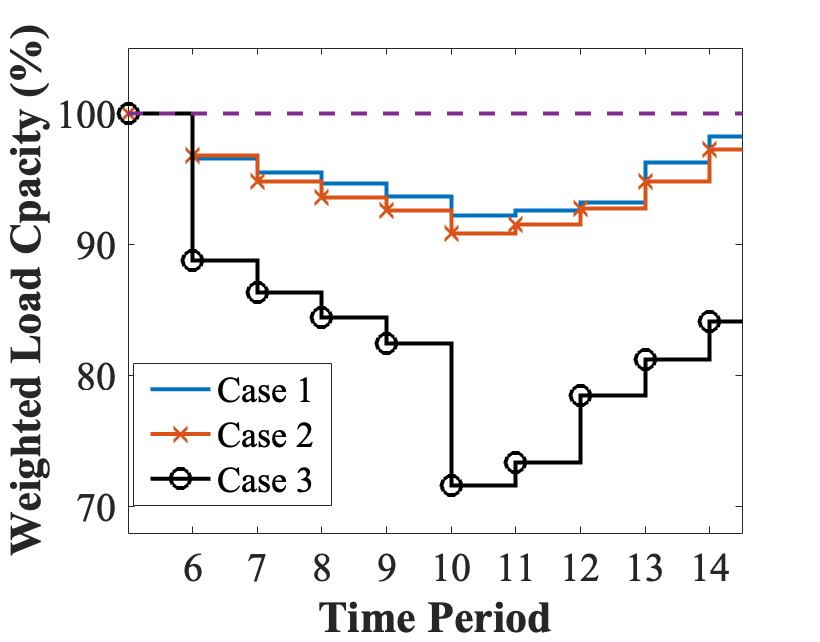}}
\subfigure[Expected weighted load shedding under different hardening budget $B^{\rm h}$]{
\includegraphics[width=6.3cm,height=4.8cm]{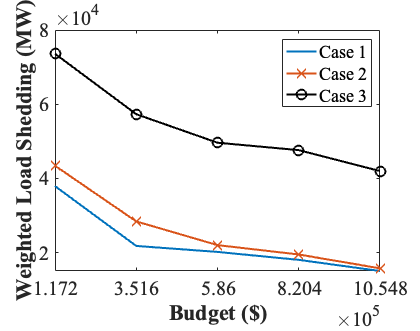}}

\caption{Performances of three cases in IEEE 33-node system}

\label{Performances}
\vspace{-2ex}
\end{figure}

Moreover, to investigate the contributions of DG allocation and proactive reconfiguration, three cases are considered: 1) Case 1: the proposed method is adopted; 2) Case 2: DGs are randomly allocated; 3) Case 3: Proactive reconfiguration is not considered.
Fig. \ref{Performances}(a) shows the expected weighted load curve under hurricane track scenario 1. The expected uninterrupted load decreases in all three cases after the hurricane landed at $t=6$, and gradually restores after $t=12$. During the period 10, the weighted loads in three cases are $92.2\%$, $90.9\%$ and $71.6\%$ of their normal value. 
Obviously, the optimal placement of DGs has substantial impacts on the performances, 
due to their capability of locally supporting critical loads without relying on networks. 

On the other hand, stress tests with regard to hardening budget is also conducted as shown in Fig. \ref{Performances}(b). In all three cases, the EWLS greatly reduces with increasing budget. Notably, the contribution of proactive reconfiguration to the EWLS decrease is $14.5\%$ when budget is $\$1.2\times 10^5$, much greater than the reduction of $4.6\%$ when budget is $\$10.5\times 10^5$, indicating the importance of initial network topology given restricted budget. Moreover, we observe in case 1 that the marginal reduction in EWLS becomes less when budget exceeds $\$5.86\times10^5$. Therefore, this plot can also aid planners in selecting a more cost-effective budget level.
\vspace{-0.5ex}
\subsection{In-Sample and Out-of-Sample Evaluations}
\begin{table}[]\scriptsize
\centering
\caption{Different prior assumptions of four strategies}
\label{4-Strategies2}
\vspace{1ex}
\begin{tabular}{lccc}
\hline \hline
      \textbf{Strategy} & \begin{tabular}[c]{@{}c@{}}\textbf{Distributional} \\ \textbf{Ambiguity}\end{tabular} & \begin{tabular}[c]{@{}c@{}}\textbf{Decision-dependency} \\ \textbf{of Distributions}\end{tabular} & \begin{tabular}[c]{@{}c@{}}\textbf{Uncertainty of} \\\textbf{Hurricane Intensity} \end{tabular} \\ \hline
\textbf{1} (DD-DRRE) & \small \checkmark                  & \small \checkmark              & \small \checkmark                       \\
\textbf{2} (DD-SRE) & \small $\times$                   & \small \checkmark             & \small \checkmark                       \\
\textbf{3} (RRE) & \small $\times$                   & \small $\times$             & \small \checkmark                       \\
\textbf{4} (DD-DRRE with FHI) & \small \checkmark                   & \small \checkmark             & \small $\times$                      \\ \hline \hline
\end{tabular}
\end{table}
To demonstrate the potential value of incorporating the DDU and distributional ambiguity, we examine four strategies derived from four counterpart models with distinct prior assumptions. According to TABLE \ref{4-Strategies2}, Strategy 1 is generated from our proposed model. Strategy 2 is derived from the SP-based DD-SRE model, 
and Strategy 3 is drawn from the RO-based RRE model. 
By setting the wind speed to its expected value, Strategy 4 eliminates the random property of hurricane intensity from the DD-DRRE model, which is denoted as the fixed hurricane intensity (FHI) strategy. 
\begin{table}\footnotesize
\centering
\caption{Comparison of WLS of four strategies in IEEE 33-node system}
\label{33-4Strategies2}
\vspace{1ex}
\begin{tabular}{cccc}
\hline\hline
\textbf{Strategy} & \textbf{OBJ} (kW) & \textbf{WCD} (kW) & \textbf{RGD} (kW) \\ \hline
\textbf{1}        & 20220.6                                                     & 19494.3                                                     & 18290.7                                                     \\
\textbf{2}        & 18094.7                                                    & 22314.8                                                     & 20167.9                                                     \\
\textbf{3}        & 40515.4                                                     & 21791.5                                                     & 22578.6                                                     \\
\textbf{4}                 & 17046.6                                                     & 32714.6                                                     & 29942.4                                                     \\ 
\hline\hline
\end{tabular}
\end{table}

The left side of TABLE \ref{33-4Strategies2} summarizes the in-sample and out-of-sample performances in the IEEE 33-node system. The second column (EXP) presents the in-sample EWLS. 
In the presence of DDU and distributional ambiguity, we test their out-of-sample EWLS using two test sets: 1) the WCD set, whose scenarios are generated using the worst-case distribution obtained in proposition 1; and 2) the RGD set, whose scenarios are obtained from random distributions within SWDD-ASs. The third and fourth columns of TABLE IV report the weighted averages under these two tests, respectively.

As shown in TABLE \ref{33-4Strategies2}, although Strategy 1 obtained by our method does not have the lowest in-sample expectation, it does have the lowest EWLS under WCD and RGD tests.
The lower in-sample EWLS in Strategy 2 than in Strategy 1 and 3 is mostly due to its optimism in the accuracy of empirical distributions, but this also results in a $14.5\%$ and a $23.4\%$ increase in results under WCD and RGD tests.
For strategy 3, the in-sample EWLS is the highest owing to the pessimistic prior assumption that the worst-case contingency scenario will occur.  
The out-of-sample performances of Strategy 4 are the worst, demonstrating the critical role of capturing the random nature of hurricane intensity.

\begin{figure}
\centering
\includegraphics[width=5cm,height=5cm]{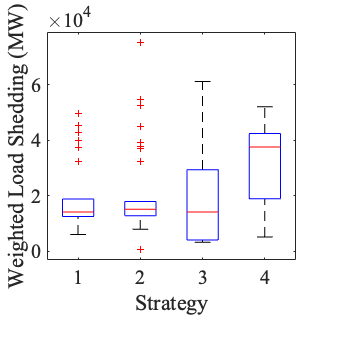}
\caption{Boxplot of WLS under WCD in IEEE 33-node system}
\label{33-boxplot2}
\end{figure}

Moreover, in order to compare the performances of four strategies under worst-case distributions, Fig. \ref{33-boxplot2} depicts a boxplot graph of the weighted load shedding (WLS) under different scenarios in the WCD set. With a comparable narrow span and positive skewness, our proposed Strategy 1 can produce relatively consistent and stable results with few outliers. While Strategy 2 has a similar pattern to Strategy 1, it has a higher median and more scattered outliers (where the median is 10.5\% higher than that of Strategy 1 and the maximal outlier equals to 79062.3 MW), indicating that it is more vulnerable to contingency misspecification.  
With regards to Strategy 3, its boxplot graph has a much larger span than those of Strategies 1 and 2, indicating that the distribution-free RO-based strategy with no DDU consideration cannot guarantee a stable result. 
Finally, the wide span and negative skewness of boxplot for the Stratey 4 demonstrate the necessity of more accurate hurricane knowledge.

\begin{figure}
\centering

\subfigure[Strategy 2 derived from the DD-SRE model]{
\includegraphics[width=12cm,height=4.4cm]{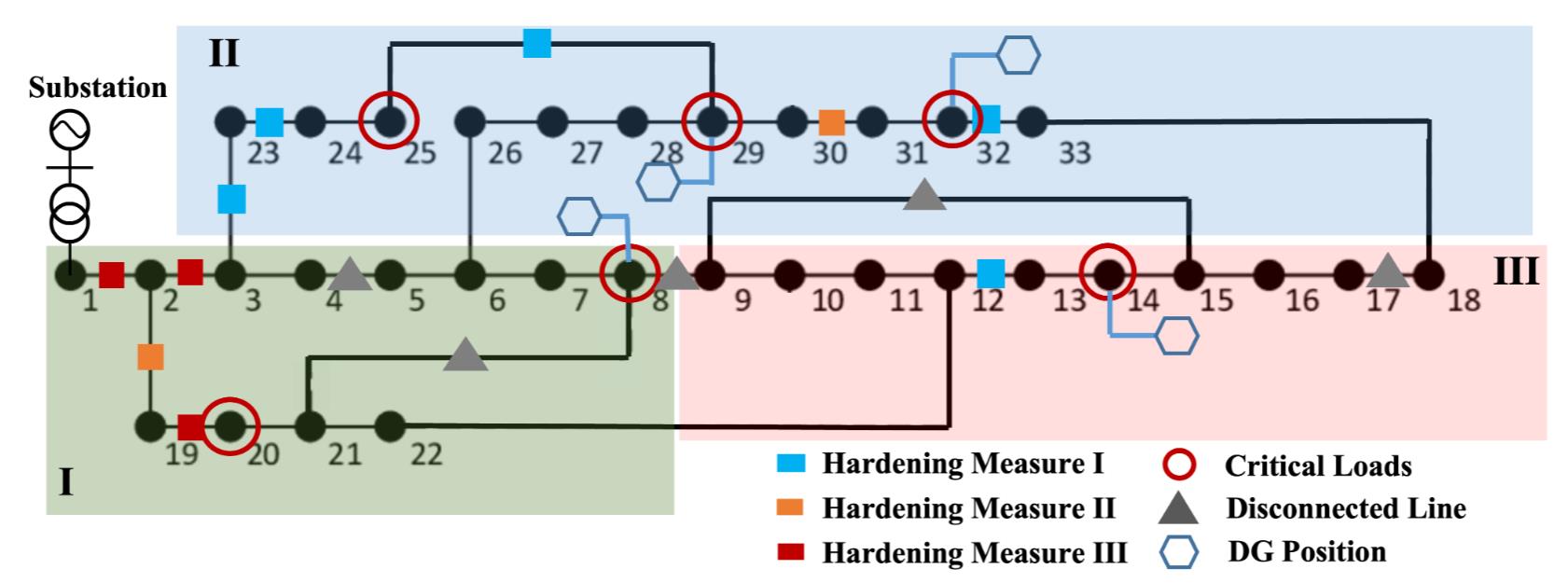}}
\subfigure[Strategy 3 derived from the RRE model]{
\includegraphics[width=12cm,height=4.4cm]{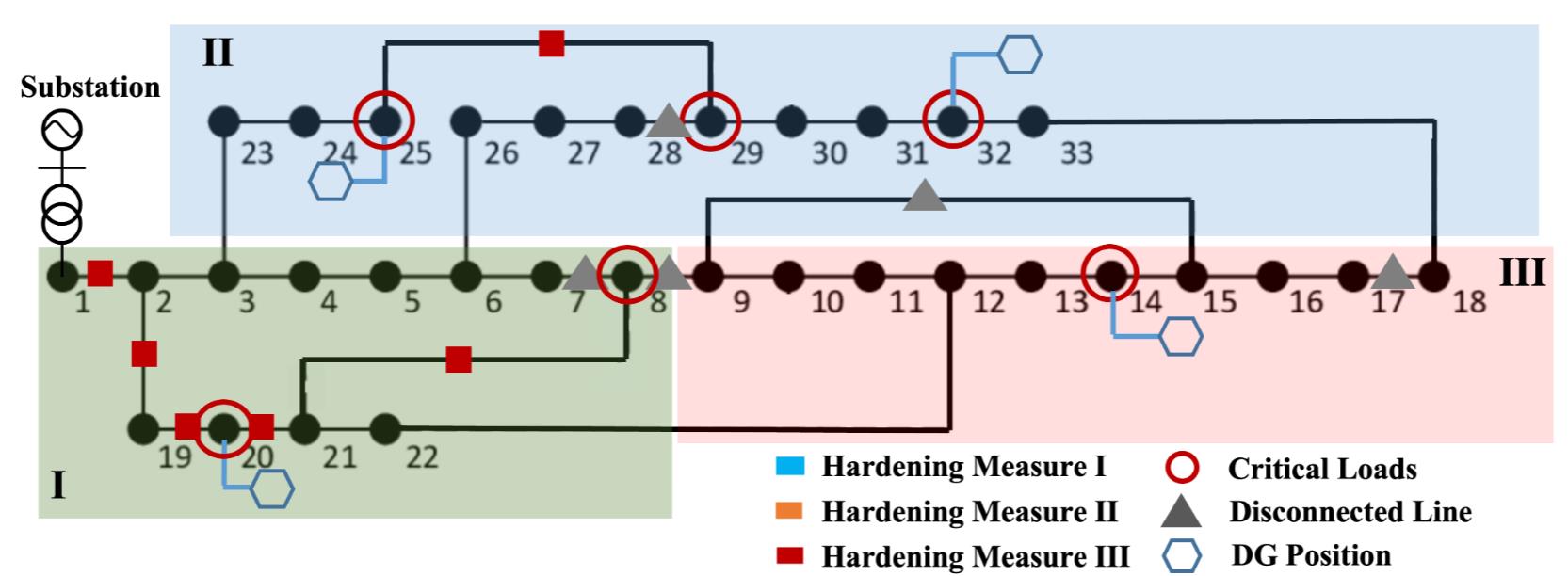}}

\caption{Optimal enhancement strategies under counterpart models}
\label{33-counterparts2}
\vspace{-2ex}
\end{figure}

\subsection{Enhancement Results of Strategy 2 and Strategy 3}
Then, we further analyze the detailed enhancement results of Strategy 2 and Strategy 3 shown in Fig. \ref{33-counterparts2}. In Strategy 2, DG allocations are the same as those in Strategy 1. However, the likely underestimation toward failure rates of line 29-30 and line 31-32 reduces the expected transmittable power from DG at node 32 to critical load node 25.
Further, optimism about the status of line 2-19 results in a less effective measure II, as well as the first line disconnection of line 9-21. These measures contribute to the degraded out-of-sample EWLS by reducing the survivability of critical load node 20 from both the substation and the DG at node 8. 
On the other hand, Strategy 3 considers only the most effective measure III as a candidate due to its conservatism. Due to the limited capacity of DG at node 25 and the comparably high failure rate of path from node 32 to 29, power supply at node 29 cannot be assured in both in-sample or out-of-sample tests. Additionally, the unrealistic assumption that hardened lines would always survive results in the unanticipated WLS in node 8 in the out-of-sample test. Owing to the insufficient hardening measures, the failure rates of noncritical nodes with a large load capacity rise, such as nodes 12 and 23 (both of which have a maximum load capacity of roughly 400kW). 


\subsection{Results of VoDA and VoMI}

\begin{figure}

\centering

\subfigure[VoDA (\%)]{
\includegraphics[width=6cm,height=4.5cm]{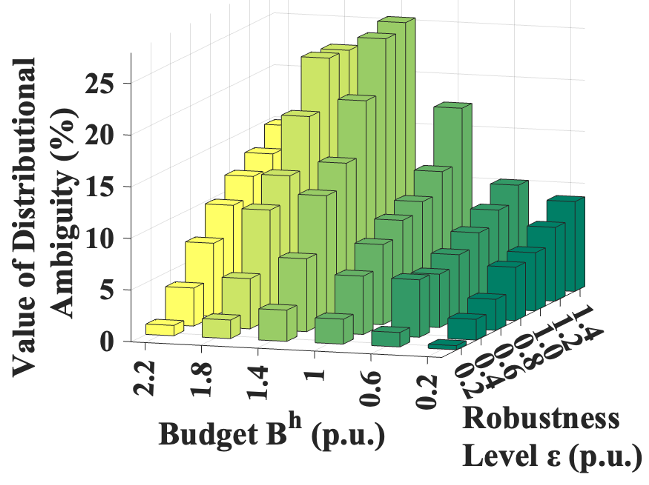}}
\subfigure[VoMI (\%)]{
\includegraphics[width=6cm,height=4.5cm]{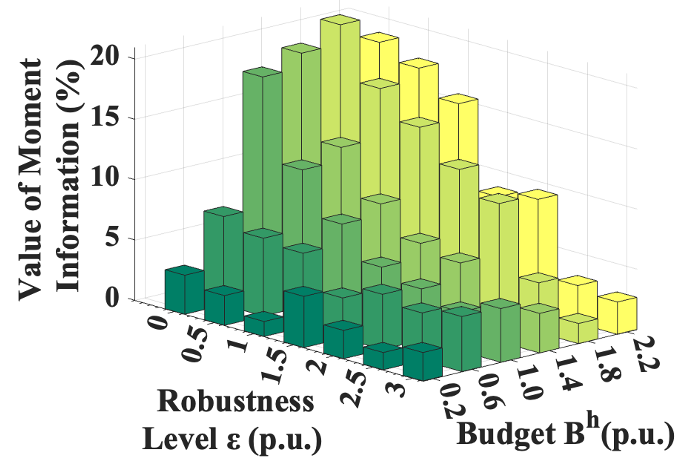}}

\caption{Expected VoDA and VoMI under various hardening budgets and robustness levels in IEEE 33-node test system}

\label{POPP2}
\vspace{-4ex}
\end{figure}

In the presence of ambiguity and DDU, the hidden risks and applicability of the prevalent SP- and RO-based counterparts are investigated by computing VoDA and VoMI under different line hardening budgets $B^{\rm L}$ and robustness levels $\boldsymbol{\varepsilon}$. The results are depicted in Fig. \ref{POPP2}, which have the following properties.

First, given any hardening budget, the VoDA represented in Fig. \ref{POPP2}(a) grows monotonically in $\boldsymbol{\varepsilon}$, reaching 26.14\% at $\boldsymbol{\varepsilon} = 1.4\ \rm p.u.$ when the budget is $\$1.4 \times 5.16 \times 10^5$. The rationale is that an increased $\boldsymbol{\varepsilon}$ implies that either the empirical fragility estimation is less trustworthy or the planner is more risk-averse. 
In this regard, simply adopting the SP-based strategy will inevitably result in greater retrospective regret. 
On the other hand, VoMI in Fig. \ref{POPP2}(b) exhibits an overall non-increasing pattern in terms of $\boldsymbol{\varepsilon}$. 
This is because mistrust in empirical contingency distributions and associated DDU will result in an increasing homogeneity between our proposed DRO-based model and the RO-based model.

Second, given a fixed $\boldsymbol{\varepsilon}$, both the VoDA and VoMI first grow with the budget $B^{\rm h}$, but subsequently decline after reaching their peaks around $1.4\ \rm p.u.$ and $1.8\ \rm p.u.$, respectively. 
The downward trends after the peak imply that more investment budgets might help smooth out the effects of model misspecification.

Finally, we also notice that there are several circumstances making the performance of our proposed strategy relatively indistinguishable from its counterparts. 
When the budget is inadequate, both the VoDA and VoMI are trivial due to the limited hardened line. 
Besides this, when empirical estimation of component failure rates is highly accurate (i.e., $\boldsymbol{\varepsilon}$ is small), the SP-based DD-SRE model is acceptable. Moreover, the RO-based RRE model is a reasonable substitute when the underlying contingency distributions and associated DDU are highly unpredictable. 
Still, 
our proposed approach is obviously superior 
in most situations with moderate budgets and risk-aversion levels.

\end{document}